\documentclass[letterpaper,10pt,final]{article}

\usepackage[left=3cm,top=2cm,right=3cm,bottom=2cm]{geometry} 
\usepackage{multicol} 
\usepackage{url}
\usepackage{cuted}
\usepackage{flushend}

\usepackage[utf8]{inputenc} 
\usepackage[english]{babel} 
\usepackage{verbatim} 
\usepackage{color} 
\usepackage{chapterbib} 
\usepackage[normalem]{ulem}

\usepackage{amsmath}
\usepackage{amsfonts}
\usepackage{amssymb}
\usepackage{bbm}
\usepackage{cancel} 
\usepackage{youngtab}

\usepackage{graphicx}
\usepackage{feynmp} 
\usepackage{subfigure} 
\usepackage{float} 

\usepackage{multirow} 

\usepackage{pgfplots}
\usepackage{xparse}

\usetikzlibrary{positioning,arrows,patterns}
\usetikzlibrary{decorations.markings}
\usetikzlibrary{calc}

\tikzset{
  photon/.style={decorate, decoration={snake}, draw=black},
  fermion/.style={draw=black, postaction={decorate},decoration={markings,mark=at position .55 with {\arrow{>}}}},
  vertex/.style={draw,shape=circle,fill=black,minimum size=3pt,inner sep=0pt},
  vacuum/.style={draw,shape=rectangle,fill=black,minimum size=3pt,inner sep=0pt},
}

\NewDocumentCommand\semiloop{O{black}mmmO{}O{above}}
{%
\draw[#1] let \p1 = ($(#3)-(#2)$) in (#3) arc (#4:({#4+180}):({0.5*veclen(\x1,\y1)})node[midway, #6] {#5};)
}

\begin{document}

\title{Neutrino and CP-even Higgs boson masses in a nonuniversal $\mathrm{U}(1)'$ extension}
\author{S.F. Mantilla\thanks{sfmantillas@unal.edu.co}, R. Martinez\thanks{remartinezm@unal.edu.co}, F. Ochoa\thanks{faochoap@unal.edu.co} }

\maketitle
\begin{center}
\textit{Departamento de F\'{i}sica, Universidad Nacional de Colombia,
Ciudad Universitaria, K. 45 No. 26-85, Bogot\'a D.C., Colombia} 
\par\end{center}



\vspace*{0cm}
\begin{center}\rule{0.9\textwidth}{0.1mm} \end{center}
\begin{abstract}
We propose a new anomaly-free and family nonuniversal $\mathrm{U}(1)'$ extension of the standard model with the addition of two scalar singlets and a new scalar doublet. The quark sector is extended by adding three exotic quark singlets, while the lepton sector includes two exotic charged lepton singlets, three right-handed neutrinos and three sterile Majorana leptons to obtain the fermionic mass spectrum of the standard model. The lepton sector also reproduces the elements of the PMNS matrix and the squared-mass differences data from neutrino oscillation experiments. Also, analytical relations of the PMNS matrix are derived via the inverse see-saw mechanism, and numerical predictions of the parameters in both normal and inverse order scheme for the mass of the phenomenological neutrinos are obtained. 
We employed a simple seesaw-like method to obtain analytical mass eigenstates of the CP-even $3\times 3$ mass matrix of the scalar sector.
 
\begin{center}\rule{0.9\textwidth}{0.1mm} \end{center}
\vspace*{0.5cm}
\end{abstract}

\section{Introduction}
Despite all its success, the Standard Model (SM) of Glashow, Weinberg and Salam  \cite{SM} 
has some unexplained features, which has motivated many models and extensions. In particular, the observed fermion mass hierarchies, their mixing and the three family structure are not explained in the SM. From the phenomenological point of view, it is possible to describe some features of the mass hierarchy by assuming zero-texture Yukawa matrices \cite{textures}. Models with spontaneously broken flavor symmetries may also produce hierarchical mass structures. For example, in models with gauge symmetry $\mathrm{\mathrm{SU}(2)}_{L}\otimes \mathrm{\mathrm{SU}(2)}_{R}\otimes \mathrm{\mathrm{U}(1)}_{B-L}$, the electroweak doublets exhibit a discrete symmetry after the spontaneous symmetry breaking, obtaining Fritzsch zero-texture mass matrices \cite{fritzsch1978} in the basis $\mathbf{U}=(u_{0},c_{0},t_{0})$ of the form:

\begin{equation}
-\left\langle \mathcal{L}_{Y,U} \right\rangle_{0} = \overline{\mathbf{U}_{L}}
\left( \begin{matrix}
0	&	a	&	0	\\	a^{*}	&	0	&	b	\\	0	&	b^{*}	&	c
\end{matrix} \right) \mathbf{U}_{R} + \mathrm{h.c.}
\end{equation}
The zero-texture of the above matrix can describe the mass spectrum in the quark sector and the CP violation phase observed in the experiments. This mass structure can also be obtained in the lepton sector, as shown by Fukugita, Tanimoto y Yanagida \cite{fty1993}, where very small mass values are predicted through a seesaw mechanism. In addition, these type of models contain Majorana neutrinos which induce matter-antimatter asymmetry through leptogenesis \cite{leptogenesis}. 


Another issue that the SM can not explain is the observation of neutrino oscillations. These observations have been confirmed by many experiments from four different sources: solar neutrinos as in Homestake \cite{homestake}, SAGE \cite{sage}, GALLEX \& GNO \cite{gallex}, SNO \cite{sno}, Borexino \cite{borexino} and Super-Kamiokande \cite{super-kamiokande} experiments, atmospheric neutrinos as in IceCube \cite{ice-cube}, neutrinos from reactors as KamLAND \cite{kamland}, CHOOZ \cite{chooz}, Palo Verde \cite{palo-verde}, Daya Bay \cite{daya-bay}, RENO \cite{reno} and SBL \cite{sbl}, and from accelerators as in MINOS \cite{minos}, T2K \cite{t2k} and NO$\nu $A \cite{nova}. The experimental data are compatible with the hypothesis that at least two species of neutrinos have mass, where the left-handed flavor neutrino fields are linear combinations of mass eigenstates

\begin{equation}
\left| \nu^{a}_{L}\right\rangle = \sum_{i=1,2,3} U_{ai} \left| \nu^{i}_{L}\right\rangle, \quad a=e,\mu,\tau
\end{equation}
where $U$ is the Pontecorvo-Maki-Nakagawa-Sakata (PMNS) matrix, which can be parameterized as function of three mixing angles and one CP violating phase \cite{neutrinodata, fritzsch2013}. However, the experiments can not determine the true nature of the active neutrinos (Majorana or Dirac) nor the absolute values of their mass. Table \ref{tab:Neutrino-data} shows the parameters from references \cite{neutrinodata, esteban-gonzalez-maltoni} and available at NuFIT 3.0 \cite{nufit}, where two hierarchies are assumed: normal ordering (NO), where the squared mass difference between the third and first species accomplish $\Delta m_{31}^2 > 0$, and inverted ordering (IO), where $\Delta m_{32}^2 < 0$ between the second and third species.  

\begin{table}
\centering
\begin{tabular}{c|c|c}
	&	Normal Ordering (NO)	&	Inverted Ordering (IO)	\\	\hline
$\sin^{2}\theta_{12}$	&	$0.308^{+0.013}_{-0.012}$	&	$0.308^{+0.013}_{-0.012}$	\\	\hline
$\sin^{2}\theta_{23}$	&	$0.440^{+0.023}_{-0.019}$	&	$0.584^{+0.018}_{-0.022}$	\\	\hline
$\sin^{2}\theta_{13}$	&	$0.02163^{+0.00074}_{-0.00074}$	&	$0.02175^{+0.00075}_{-0.00074}$	\\	\hline
$\delta_{\mathrm{CP}}$	&	$289^{+38}_{-51}$	&	$269^{+39}_{-45}$	\\ \hline
$\dfrac{\Delta m_{21}^{2}}  {10^{-5}\mathrm{\,eV^{2}}}$	&	$7.49^{+0.19}_{-0.17}$	&	$7.49^{+0.19}_{-0.17}$	\\ \hline
$\dfrac{\Delta m_{3\ell}^{2}}  {10^{-3}\mathrm{\,eV^{2}}}$	&	$+2.526^{+0.039}_{-0.037}$	&	$-2.518^{+0.038}_{-0.037}$	\\\hline
\end{tabular}
\caption{Three-flavor oscillation parameter values at 1$\sigma$ reported by \cite{esteban-gonzalez-maltoni,neutrinodata}. $\ell=1$ for NO and $2$ for IO.}
\label{tab:Neutrino-data}
\end{table}

On the other hand, in order to obtain tiny neutrino masses, two method can be used: radiative corrections and see-saw mechanism. 
The latter scheme has been studied in the literature and is considered as one of the most traditional schemes for the explanation of smallness of neutrino masses. The see-saw mechanism implies the addition of a lepton number-violating high energy scale ($M$), which gives masses to light neutrinos as $m_\nu=v_w^2/M$. There are some basic ways to implement this mechanism: a heavy right-handed Majorana neutrino ${\nu_{R}}$ mixed to the corresponding left-handed neutrino $\nu_L$ via the SM scalar doublet (type I seesaw), a heavy scalar triplet bosons (type II), or a heavy fermionic triplet (type III). Since the new scale $M$ associated with the new fields is high ($\sim 10^{12}$ GeV), this mechanism has the problem that is not accessible to be test in experiments. However, there is another possibility: the inverse see-saw mechanism (ISS), where a very light Majorana neutrino ${N_{R}}$ is incorporated, such that in the basis $(\nu_{L},{\nu_{R}}^{c},{N_{R}}^{c})$ the mass matrix has the form of Fristzsch zero-texture:
\begin{equation}
\mathcal{M}_{\nu} = \left( \begin{matrix}
0	&	m_{\nu}^{\mathrm{T}}	&	0	\\	m_{\nu}	&	0	&	m_{N}^{\mathrm{T}}	\\	0	&	m_{N}	&	M_{N}
\end{matrix} \right).
\end{equation}
where the submatrix $m_{N}$ has components of the order of the TeV scale, while $M_N$ is of the order of the KeV scale, in order to obtain active neutrinos at the sub-eV scale. The inverse see-saw mechanism was proposed in \cite{inverseseesaw}. This mechanism has also been implemented in $\mathrm{\mathrm{SU}(3)}_{L}\otimes\mathrm{\mathrm{U}(1)}_{X}$ models in order to study the $\mu\rightarrow e\gamma$ decay \cite{catano2012}.


On the other hand, the discovery of the Higgs boson at ATLAS\cite{Aad:2012tfa} and CMS \cite{Chatrchyan:2012ufa} whose mass is $125$ GeV opens the window to propose other scalar fields. A new scalar sector is considered as extension to the SM in order to explain some phenomenological aspects. One of the most studied SM extension is the two-Higgs-doublet-model (2HDM) which proposes the existence of two scalar doublets whose scalar potential mixes them together obtaining two charged scalar bosons $H^{\pm}$, a CP-odd pseudoscalar $A^{0}$ and two CP-even scalar bosons $h$ and $H$ \cite{higgshunter}. This model was motivated in order to give masses to up-like and down-like quarks \cite{2HDM-review} where vacuum expectation values (VEV) $v_{2}$ and $v_{1}$ are related to the electroweak VEV by $v^{2}=v_{2}^{2}+v_{1}^{2}$. 

There are also extensions to the 2HDM adding a new scalar singlet $\chi$, as in the Next-to-Minimal 2HDM (N2HDM) \cite{He:2008qm}. In some cases, this additional singlet implement the spontaneous symmetry breaking (SSB) of an additional $U(1)^\prime$ gauge symmetry through the acquisition of non-vanishing VEV $v_{\chi}$, and consequently its imaginary part become in the would-be Goldstone boson eaten by the corresponding gauge boson of $U(1)^\prime$ \cite{somepheno}. Furthermore, if this SSB happens at a higher scale than the electroweak ($v\ll v_{\chi}$), the CP-even mass matrix exhibits an internal hierarchy which allows us to employ a perturbative see-saw-like method in order to obtain analytical expressions for the mass eigenvalues and angles of the corresponding mixing matrix.

Models with extra $\mathrm{U}(1)'$ symmetry are one of the most studied extensions of the SM, which implies many phenomenological and theoretical advantages including flavor physics \cite{flavor}, neutrino physics \cite{neutrino}, dark matter \cite{DM}, among other effects \cite{moretti}. A complete review of the above possibilities can be found in reference \cite{review}. In particular, family non-universal $\mathrm{U}(1)'$ symmetry models have many well-established motivations. For example, they provide hints for solving the SM flavor puzzle, where even though all the fermions acquire masses at the same scale, $\upsilon =246$ GeV, experimentally they exhibit very different mass values. These models also imply a new $Z'$ neutral boson, which contains a large number of phenomenological consequences at low and high energies \cite{zprime-review}. In addition to the new neutral gauge boson $Z'$, an extended fermion spectrum is necessary in order to obtain an anomaly-free theory. Also, the new symmetry requires an extended scalar sector in order to (i) generate the breaking of the new Abelian symmetry and (ii) obtain heavy masses for the new $Z'$ gauge boson and the extra fermion content. A non-universal $\mathrm{U}(1)'$ model in the quark sector was proposed in \cite{somepheno}, obtaining zero-texture quark mass matrices with hierarchical structures, where three quarks [up, down and strange] acquire masses at the MeV scale, and three quarks [charm, bottom and top] exhibit masses at the GeV scale. Additional phenomenological consequences of this model were studied in \cite{DM-martinez-I, DM-martinez-II, DM-jhep} including effects on scalar DM. 

The main purpose of this paper is to construct an anomaly-free and family non-universal $\mathrm{U}(1)'$ symmetry model in both the quark and leptonic sector, with extra lepton and quark singlets, two scalar doublets, and two scalar singlets. The leptonic sector includes new charged and right-handed neutral leptons, and sterile Majorana neutrinos in order to reproduce the PMNS matrix and the observed mass structure of the leptons. In Sec. \ref{sect:Model}, we describe the spectrum and most important properties of the model. We also show the scalar and gauge Lagrangians, including rotations into mass eigenvectors. In Sec. \ref{sect:Fermion-masses} we show how mass structures in the fermion sector are predicted in the model, first for the quark sector in subsection \ref{subsect:Quark-masses}, and later for the leptonic sector in subsection \ref{subsect:Lepton-masses}. Sec. \ref{sect:PMNS-matrix} is devoted to obtain some phenomenological parameters from neutrino oscillation data at 1 $\sigma$. Finally, the Sec. \ref{sect:Conclusions} outlines the main results of the article.

\section{Non universal  model with extra $\mathrm{\mathrm{U}(1)}_{X}$ symmetry}
\label{sect:Model}

The model proposes the existence of a new non-universal gauge group $U(1)'$ whose gauge boson and coupling constant are $Z_{\mu}'$ and $g_{X}$, respectively. It brings the following triangle anomaly equations:

\begin{eqnarray}
\left[\mathrm{\mathrm{SU}(3)}_{C} \right]^{2} \mathrm{\mathrm{U}(1)}_{X} \rightarrow & A_{C} =& \sum_{Q}X_{Q_{L}} - \sum_{Q}X_{Q_{R}},	\\
\left[\mathrm{\mathrm{SU}(2)}_{L} \right]^{2} \mathrm{\mathrm{U}(1)}_{X} \rightarrow & A_{L}  =& \sum_{\ell}X_{\ell_{L}} + 3\sum_{Q}X_{Q_{L}},	\\
\left[\mathrm{\mathrm{U}(1)}_{Y} \right]^{2}   \mathrm{\mathrm{U}(1)}_{X} \rightarrow & A_{Y^{2}}=&
	\sum_{\ell, Q}\left[Y_{\ell_{L}}^{2}X_{\ell_{L}}+3Y_{Q_{L}}^{2}X_{Q_{L}} \right] - \sum_{\ell,Q}\left[Y_{\ell_{R}}^{2}X_{L_{R}}+3Y_{Q_{R}}^{2}X_{Q_{R}} \right],	\\
\mathrm{\mathrm{U}(1)}_{Y}   \left[\mathrm{\mathrm{U}(1)}_{X} \right]^{2} \rightarrow & A_{Y}=&
	\sum_{\ell, Q}\left[Y_{\ell_{L}}X_{\ell_{L}}^{2}+3Y_{Q_{L}}X_{Q_{L}}^{2} \right] - \sum_{\ell, Q}\left[Y_{\ell_{R}}X_{\ell_{R}}^{2}+3Y_{Q_{R}}X_{Q_{R}}^{2} \right],	\\
\left[\mathrm{\mathrm{U}(1)}_{X} \right]^{3} \rightarrow & A_{X}=&
	\sum_{\ell, Q}\left[X_{\ell_{L}}^{3}+3X_{Q_{L}}^{3} \right] - \sum_{\ell, Q}\left[X_{\ell_{R}}^{3}+3X_{Q_{R}}^{3} \right],		\\
\left[\mathrm{Grav} \right]^{2}   \mathrm{\mathrm{U}(1)}_{X} \rightarrow & A_{\mathrm{G}}=&
	\sum_{\ell, Q}\left[X_{\ell_{L}}+3X_{Q_{L}} \right] - \sum_{\ell, Q}\left[X_{\ell_{R}}+3X_{Q_{R}} \right],
\end{eqnarray}
where the sums in $Q$ run over quarks 
while $\ell $ runs over leptons with nontrivial $\mathrm{U}(1)_X$ values. 
$Y$ is the corresponding  weak hypercharge. The fermion content compatible with the above conditions is composed by ordinary SM particles but also new exotic non-SM particles, as shown in table \ref{tab:Fermionic-content}, where column $X$ contains the quantum numbers of the extra $\mathrm{U}(1)_X$ and the $\mathbf{Z}_{2}$ column presents their corresponding $Z_2$-parity under a new $Z_2$ discrete symmetry. Some properties of this spectrum are outlined below:

\begin{table}
\centering
\begin{tabular}{cccc|cccc|}
\hline\hline
Quarks	&	$X$	&$\mathbf{Z}_{2}$&&	Leptons	&	$X$&$\mathbf{Z}_{2}$	\\ \hline 
\multicolumn{7}{c}{SM Fermionic Isospin Doublets}	\\ \hline\hline
$q^{1}_{L}=\left(\begin{array}{c}U^{1} \\ D^{1} \end{array}\right)_{L}$
	&	$+1/3$	&$+$	&&
$\ell^{e}_{L}=\left(\begin{array}{c}\nu^{e} \\ e^{e} \end{array}\right)_{L}$
	&	$0$	&$+$	\\
$q^{2}_{L}=\left(\begin{array}{c}U^{2} \\ D^{2} \end{array}\right)_{L}$
	&	$0$	&$-$	&&
$\ell^{\mu}_{L}=\left(\begin{array}{c}\nu^{\mu} \\ e^{\mu} \end{array}\right)_{L}$
	&	$0$	&$+$		\\
$q^{3}_{L}=\left(\begin{array}{c}U^{3} \\ D^{3} \end{array}\right)_{L}$
	&	$0$	&$+$	&&
$\ell^{\tau}_{L}=\left(\begin{array}{c}\nu^{\tau} \\ e^{\tau} \end{array}\right)_{L}$
	&	$-1$	&$+$	\\   \hline\hline

\multicolumn{7}{c}{SM Fermionic Isospin Singlets}	\\ \hline\hline
\begin{tabular}{c}$U_{R}^{1,3}$\\$U_{R}^{2}$\\$D_{R}^{1,2,3}$\end{tabular}	&	 
\begin{tabular}{c}$+2/3$\\$+2/3$\\$-1/3$\end{tabular}	&
\begin{tabular}{c}$+$\\$-$\\$-$\end{tabular}	&&
\begin{tabular}{c}$e_{R}^{e,\tau}$\\$e_{R}^{\mu}$\end{tabular}	&	
\begin{tabular}{c}$-4/3$\\$-1/3$\end{tabular}	&	
\begin{tabular}{c}$-$\\$-$\end{tabular}\\   \hline \hline 

\multicolumn{3}{c}{Non-SM Quarks}	&&	\multicolumn{3}{c}{Non-SM Leptons}	\\ \hline \hline
\begin{tabular}{c}$T_{L}$\\$T_{R}$\end{tabular}	&
\begin{tabular}{c}$+1/3$\\$+2/3$\end{tabular}	&
\begin{tabular}{c}$-$\\$-$\end{tabular}	&&
\begin{tabular}{c}$\nu_{R}^{e,\mu,\tau}$\\$N_{R}^{e,\mu,\tau}$\end{tabular} 	&	
\begin{tabular}{c}$1/3$\\$0$\end{tabular}	&	
\begin{tabular}{c}$-$\\$-$\end{tabular}\\
$J^{1,2}_{L}$	&	  $0$ 	&$+$	&&	$E_{L},\mathcal{E}_{R}$	&	$-1$	&$+$	\\
$J^{1,2}_{R}$	&	 $-1/3$	&$+$	&&	$\mathcal{E}_{L},E_{R}$	&	$-2/3$	&$+$	\\ \hline \hline
\end{tabular}
\caption{Non-universal $X$ quantum number and $\mathbf{Z}_{2}$ parity for SM and non-SM fermions.}
\label{tab:Fermionic-content}
\end{table}

\begin{enumerate}

\item The $\mathrm{U}(1)_X$ symmetry is only non-universal in the left-handed SM quark sector: the first family $1$ has $X=1/3$ while the last two $2,3$ have $X=0$. Leptons exhibit non-universal charges in both left- and right-handed sectors: $X=0$ for the left-handed components $e,\mu$ and $X=-1$ for $\tau$, while for the right-handed components $X=-4/3$ for $e,\tau$ and $X=-1/3$ for $\mu$. We use the following assignation for the phenomenological families:
\begin{equation}
U^{1,2,3}=(u,c,t), \hspace{0.5cm} D^{1,2,3}=(d,s,b), 
\hspace{0.5cm} e^{e,\mu,\tau}=(e,\mu ,\tau ), \hspace{0.5cm} \nu^{e,\mu,\tau}=(\nu^e,\nu^\mu ,\nu^\tau ).
\label{fermion-assignation}
\end{equation}

\item In order to ensure cancellation of the gauge chiral anomalies, the model includes extra isospin singlets. The quark sector has an up $T$ and two down $J^{1,2}$ quarks. For the lepton sector, three right-handed neutrinos $\nu ^{e,\mu,\tau }_R$ and two charged leptons $E$ and $\mathcal{E}$ are added with non-trivial $U(1)_X$ charges, as shown in Tab. \ref{tab:Fermionic-content}.

\item The most natural way to obtain massive neutrinos, according to neutrino oscillations, is through a see-saw mechanism, which requires the introduction of extra Majorana neutrinos. Thus, for obtaining a realistic model compatible with massive neutrinos, three sterile Majorana neutrinos $N ^{e,\mu,\tau }_R$ are included. 
\end{enumerate}

The scalar sector of the model is shown in table \ref{tab:Scalar-content}, which exhibits the following properties:
\begin{table}
\centering
\begin{tabular}{ccc}
Scalar bosons	&	$X$	&	$\mathbf{Z}_{2}$	\\ \hline 
\multicolumn{2}{c}{Higgs Doublets}\\ \hline\hline
$\phi_{1}=\left(\begin{array}{c}
\phi_{1}^{+} \\ \dfrac{h_{1}+v_{1}+i\eta_{1}}{\sqrt{2}}
\end{array}\right)$	&	$2/3$	&	$+$	\\
$\phi_{2}=\left(\begin{array}{c}
 \phi_{2}^{+} \\ \dfrac{h_{2}+v_{2}+i\eta_{2}}{\sqrt{2}}
\end{array}\right)$	&	$1/3$	&	$-$	\\   \hline\hline
\multicolumn{2}{c}{Higgs Singlets}\\ \hline\hline
$\chi  =\dfrac{\xi_{\chi}  +v_{\chi}  +i\zeta_{\chi}}{\sqrt{2}}$	& $-1/3$	&	$+$	\\   
$\sigma$	& $-1/3$	&	$-$	\\   \hline \hline 
\end{tabular}
\caption{Non-universal $X$ quantum number for Higgs fields.}
\label{tab:Scalar-content}
\end{table}

\begin{enumerate}
\item Two scalar doublets $\phi _{1,2}$ are included with $\mathrm{U}(1)_X$ charges $+2/3$ and $+1/3$ respectively, whose vacuum expectation values (VEVs) are related to the electroweak VEV by $v = \sqrt{v _{1}^2+v _{2}^2}$. The internal $\mathbf{Z}_{2}$ symmetry is introduced in order to obtain adequate zero texture matrices.

\item An extra scalar singlet $\chi $ with VEV $\upsilon _{\chi}$ is required for the SSB of $\mathrm{U}(1)_X$ and also to generate masses to exotic isospin singlets. We assume that it happens at a larger scale $\upsilon _{\chi} \gg \upsilon$ than electroweak.

\item Another scalar singlet $\sigma $ is introduced. Since it is not essential for the symmetry breaking mechanisms, we may choose $\upsilon _{\sigma } = 0$ for its VEV.  
\end{enumerate}

Finally, in the vector sector, an extra gauge boson $Z'_{\mu}$ is required to obtain a local $\mathrm{U}(1)_X$ symmetry. The covariant derivative of the model is
\begin{eqnarray}
\label{covariant}
D_{\mu}=\partial_{\mu} - i g W_{\mu}^{\alpha} T_{\alpha}-ig'\frac{Y}{2}B_{\mu}-ig_{X} X Z'_{\mu},
\end{eqnarray}
where $2T^{\alpha}$ corresponds to the Pauli matrices for isospin doublets and $T^{\alpha}=0$ for isospin singlets. The electric charge is defined by the Gell-Mann-Nishijima relation:
\begin{equation}
Q = T_{3} + \frac{Y}{2}.
\end{equation}

\subsection{Scalar masses}
\label{subsect:Scalar-masses}
The scalar potential of the model is
\begin{equation}
\begin{split}	
V &= \mu_{1}^{2}\phi_{1}^{\dagger}\phi_{1} + \mu_{2}^{2}\phi_{2}^{\dagger}\phi_{2} + \mu_{\chi}^{2}\chi^{*}\chi + \mu_{\sigma}^{2}\sigma^{*}\sigma			\\	&
 + \frac{f}{\sqrt{2}}\left(\phi_{1}^{\dagger}\phi_{2}\chi ^{*} + \mathrm{h.c.} \right) + \frac{f'}{\sqrt{2}}\left(\phi_{1}^{\dagger}\phi_{2}\sigma ^{*} + \mathrm{h.c.} \right) \\ &
 + \lambda_{1}\left(\phi_{1}^{\dagger}\phi_{1}\right)^{2} + \lambda_{2}\left(\phi_{2}^{\dagger}\phi_{2}\right)^{2} 
 + \lambda_{3}\left(\chi^{*}\chi \right)^{2} + \lambda_{4}\left(\sigma^{*}\sigma \right)^{2} 		\\	&
 + \lambda_{5}\left(\phi_{1}^{\dagger}\phi_{1}\right) \left(\phi_{2}^{\dagger}\phi_{2}\right)
 + \lambda'_{5}\left(\phi_{1}^{\dagger}\phi_{2}\right)\left(\phi_{2}^{\dagger}\phi_{1}\right)	\\ &
 + \left(\phi_{1}^{\dagger}\phi_{1}\right)\left[ \lambda_{6}\left(\chi^{*}\chi \right) + \lambda'_{6}\left(\sigma^{*}\sigma \right) 
 \right] 		\\	&
 + \left(\phi_{2}^{\dagger}\phi_{2}\right)\left[ \lambda_{7}\left(\chi^{*}\chi \right) + \lambda'_{7}\left(\sigma^{*}\sigma \right) 
 \right] 		\\	&
 + \lambda_{8}\left(\chi^{*}\chi \right)\left(\sigma^{*}\sigma \right) + \lambda'_{8}\left[\left(\chi^{*}\sigma \right)\left(\chi^{*}\sigma \right) + \mathrm{h.c.} \right].
\end{split}
\end{equation}

After symmetry breaking the mass matrices for the scalar sector are found. For the charged scalar bosons the mass matrix  is obtained in the basis $(\phi^{\pm}_{1},\phi^{\pm}_{2})$ 

\begin{equation}
\mathit{M}_{\mathrm{C}}^{2} = \frac{1}{4}
\begin{pmatrix}
	-f\dfrac{v_{\chi}v_{2}}{v_{1}}-\lambda_{5}'{{v_{2}}^{2}} & 
	 fv_{\chi}+\lambda_{5}'v_{1}v_{2}		\\
	 fv_{\chi}+\lambda_{5}'v_{1}v_{2}		&
	-f\dfrac{v_{\chi}v_{1}}{v_{2}}-\lambda_{5}'{{v_{1}}^{2}}
\end{pmatrix}
\end{equation}
which is diagonalized by 

\begin{equation}
\mathit{R}_{\mathrm{C}} = 
\begin{pmatrix}
c_{\beta}	&	s_{\beta}	\\	-s_{\beta}	&	c_{\beta}
\end{pmatrix},
\end{equation}
where $\tan{\beta}=s_{\beta}/c_{\beta}=v_{1}/v_{2}$. The mass matrix has the eigenvalues 

\begin{equation}
\begin{split}
m_{G_{W}^{\pm}}^{2} &= 0,	\\
m_{H^{\pm}}^{2} &= -\frac{1}{4}\frac{f v_{\chi}}{s_{\beta}c_{\beta}} -\frac{1}{4}\lambda_{5}' v^2,
\end{split}
\end{equation}
yielding the Goldstone bosons $G_{W}^{\pm}$, which provide the mass to the physical $W_{\mu}^{\pm}$ gauge bosons, and two physical charged Higgs bosons $H^{\pm}$.

Regarding to the neutral scalar sector, the mass matrix of the CP-odd sector in the basis $(\eta_{1},\eta_{2},\zeta_{\chi})$ is:

\begin{equation}
\mathit{M}_{\mathrm{I}}^{2} = -\frac{f}{4}
\begin{pmatrix}
\dfrac{ {v_{2}}\, {v_{\chi}}}{ {v_{1}}} & - {v_{\chi}} &  {v_{2}}\\
- {v_{\chi}} & \dfrac{ {v_{1}}\, {v_{\chi}}}{ {v_{2}}} & - {v_{1}}\\
 {v_{2}} & - {v_{1}} & \dfrac{ {v_{1}}\, {v_{2}}}{ {v_{\chi}}}
\end{pmatrix},
\end{equation}
which can be diagonalized by the following transformation

\begin{equation}
\mathit{R}_{\mathrm{I}} = \begin{pmatrix}
c_{\beta}	&	s_{\beta}	&	0	\\
-s_{\beta}	&	c_{\beta}	&	0	\\
0	&	0	&	1
\end{pmatrix}
\begin{pmatrix}
c_{\gamma}	&	0	&	s_{\gamma}	\\
0	&	 1	&	0	\\
-s_{\gamma}	&	0	&	 c_{\gamma}
\end{pmatrix},
\end{equation}
where $\gamma$ describes the doublet-singlet mixing $\tan{\gamma} = s_{\gamma}/c_{\gamma} = {v_{\chi}}/{v s_{\beta}c_{\beta}}$.
When  $\mathit{R}_{\mathrm{I}}$ acts on $\mathit{M}_{\mathrm{I}}^{2}$ the following eigenvalues are obtained
\begin{equation}
\begin{split}
m_{G_{Z}^{0}}^{2}	&=	0,	\\
m_{G_{Z'}^{0}}^{2}	&=	0,	\\
m_{A^{0}}^{2}	&=	-\frac{1}{4}\frac{f v_{\chi}}{s_{\beta}c_{\beta}s_{\gamma}^{2}},
\end{split}
\end{equation}
where the first two are the would-be Goldstone bosons of the neutral vector bosons $Z_{\mu}$ and $Z'_{\mu}$, respectively, while the latter is a physical CP-odd pseudoscalar boson $A^{0}$. 

On the other hand, the CP-even scalar mass matrix is
\begin{equation}
\label{eq:Mass-matrix-scalar-real}
\mathit{M}_{\mathrm{R}}^{2} = 
\begin{pmatrix}
	\lambda_{1} {v_{1}^{2}}-\dfrac{1}{4}\dfrac{f v_{\chi} v_{2}}{v_{1}} &
	\hat{\lambda}_{5} { v_{1} v_{2}}+\dfrac{1}{4}{f v_{\chi}} &
	\dfrac{1}{4}\lambda_{6}{ v_{1} v_{\chi}}+\dfrac{1}{4}{f v_{2}}		\\
	\hat{\lambda}_{5} { v_{1} v_{2}}+\dfrac{1}{4}{f v_{\chi}} &
	\lambda_{2} {v_{2}^{2}}-\dfrac{1}{4}\dfrac{f v_{\chi} v_{1}}{v_{2}} & 
	\dfrac{1}{4}\lambda_{7}{ v_{2} v_{\chi}}+\dfrac{1}{4}{f v_{1}}		\\
	\dfrac{1}{4}\lambda_{6}{ v_{1} v_{\chi}}+\dfrac{1}{4}{f v_{2}}	& 
	\dfrac{1}{4}\lambda_{7}{ v_{2} v_{\chi}}+\dfrac{1}{4}{f v_{1}} & 
	\lambda_{3} {v_{\chi}^{2}}-\dfrac{1}{4}\dfrac{f v_{1} v_{2}}{v_{\chi}}
\end{pmatrix},
\end{equation}
where $\hat{\lambda}_{5}=\left(\lambda_{5}+\lambda'_{5}\right)/2$. Since this matrix exhibits a characteristic third order polynomial with non-trivial eigenvalues, it is convenient to use another approximation in order to obtain the eigenvalues and mixing angles. We propose a seesaw-like mechanism by assuming a hierarchy of VEVs through the condition $ |f|\upsilon _{\chi}, \upsilon _{\chi} ^2 \gg \upsilon ^2 $ in the matrix elements. Thus, the matrix \eqref{eq:Mass-matrix-scalar-real} can be written in blocks as
\begin{equation}
\label{eq:Mass-matrix-scalar-real-blocks}
\mathit{M}_{\mathrm{R}}^{2} = 
\begin{pmatrix}
{\mathcal{M}_{1}}	&	{\mathcal{M}_{12}^{\mathrm{T}} }	\\
{\mathcal{M}_{12}}	&	{\mathcal{M}_{2}}
\end{pmatrix},
\end{equation}
where

\begin{eqnarray}
{\mathcal{M}_{1}} &=& 
\begin{pmatrix}
	\lambda_{1} {v_{1}^{2}}-\dfrac{1}{4}\dfrac{f v_{\chi} v_{2}}{v_{1}} &
	\hat{\lambda}_{5} { v_{1} v_{2}}+\dfrac{1}{4}{f v_{\chi}}\\
	\hat{\lambda}_{5} { v_{1} v_{2}}+\dfrac{1}{4}{f v_{\chi}} &
	\lambda_{2} {v_{2}^{2}}-\dfrac{1}{4}\dfrac{f v_{\chi} v_{1}}{v_{2}}
\end{pmatrix}, \nonumber	\\
{\mathcal{M}_{12}^{\mathrm{T}}}& =& 
\begin{pmatrix}
	\dfrac{\lambda_{6}{ v_{1} v_{\chi}}}{4}+\dfrac{{f v_{2}}}{4}		\\
	\dfrac{\lambda_{7}{ v_{2} v_{\chi}}}{4}+\dfrac{{f v_{1}}}{4}
\end{pmatrix}
\approx\begin{pmatrix}
	\dfrac{\lambda_{6}{ v_{1} v_{\chi}}}{4}	\\
	\dfrac{\lambda_{7}{ v_{2} v_{\chi}}}{4}
\end{pmatrix},\nonumber	\\
{\mathcal{M}_{2}} &=& \lambda_{3} {v_{\chi}^{2}}-\dfrac{1}{4}\dfrac{f v_{1} v_{2}}{v_{\chi}}
\approx \lambda_{3} {v_{\chi}^{2}}.
\end{eqnarray}

According to the block diagonalization procedure shown in Appendix \ref{app:block}, the mass matrix \eqref{eq:Mass-matrix-scalar-real-blocks} can be decoupled into two independent blocks through a unitary transformation as:

\begin{equation}
\mathit{R}_{\mathrm{S}}^{\mathrm{T}} \mathit{M}_{\mathrm{R}}^{2} \mathit{R}_{\mathrm{S}} = 
\begin{pmatrix}
M^{2}_{hH}	&	0	\\	0	&	m_{H_{\chi}}^{2}
\end{pmatrix}.
\end{equation}
where the transformation matrix can be approximately written as

\begin{equation}
\mathit{R}_{\mathrm{S}} = 
\begin{pmatrix}
1	&	{F_{\mathrm{R}}^{T}}	\\
-{F_{\mathrm{R}}}	&	1
\end{pmatrix},
\end{equation}
with

\begin{align}
{F_{\mathrm{R}}} &\approx {\mathcal{M}_{2}}^{-1}{\mathcal{M}_{12}}	,\nonumber\\
m_{H_{\chi}}^{2} &\approx {\mathcal{M}_{2}} = \lambda_{3} {v_{\chi}^{2}},	\nonumber\\
M^{2}_{hH} &\approx {\mathcal{M}_{1}} - {\mathcal{M}_{12}^{\mathrm{T}}} {\mathcal{M}_{2}}^{-1} {\mathcal{M}_{12}^{\mathrm{T}}},
\end{align}
and

\begin{align}
\label{eq:Mass-matrix-scalar-real-block-2-times-2}
M^{2}_{hH} &=
\begin{pmatrix}
	\tilde{\lambda}_{1} {v^{2}}s_{\beta}^{2}-\dfrac{1}{4}\dfrac{f v_{\chi} }{t_{\beta}} &
	\tilde{\lambda}_{5} {v^{2}}s_{\beta}^{2}c_{\beta}^{2}+\dfrac{1}{4}{f v_{\chi}}		\\
	\tilde{\lambda}_{5} {v^{2}}s_{\beta}^{2}c_{\beta}^{2}+\dfrac{1}{4}{f v_{\chi}}		&
	\tilde{\lambda}_{2} {v^{2}}c_{\beta}^{2}-\dfrac{1}{4}f v_{\chi} t_{\beta}
\end{pmatrix}.
\end{align}
where the new tilde constants are
\begin{equation}
\begin{split}
\tilde{\lambda}_{1} &= \lambda_{1}-\frac{\lambda_{6}^2}{4 \lambda_{3}}-\frac{\lambda_{7}^2}{4 \lambda_{3} t_{\beta}^2},\\
\tilde{\lambda}_{2} &= \lambda_{2}-\frac{\lambda_{6}^2 t_{\beta}^2}{4 \lambda_{3}}-\frac{\lambda_{7}^2}{4 \lambda_{3}},\\
\tilde{\lambda}_{5} &= \hat{\lambda}_{5}-\frac{\lambda_{6}^2 t_{\beta}}{2 \lambda_{3}}-\frac{\lambda_{7}^2}{2 \lambda_{3} t_{\beta}}.
\end{split}
\end{equation}

In order to obtain the largest eigenvalue of $M^{2}_{hH} $, we neglect non-dominant terms from the condition that $f v_{\chi}\gg v_{2}^{2}, v_{1}^{2},v_{2}v_{1}$, which leads us to

\begin{align}
M^{2}_{hH} \approx -\dfrac{1}{4}f v_{\chi} 
\begin{pmatrix}
	\cot \beta &
	-1		\\	-1		&
	\tan \beta
\end{pmatrix}.
\end{align}
Due to this approximation, the new matrix has null determinant and its trace is of the order of the largest eigenvalue

\begin{equation}
m_{H}^{2} \approx \mathrm{Tr}\left[ M^{2}_{hH} \right] \approx -\frac{1}{4}\frac{f v_{\chi}}{s_{\beta}c_{\beta}}.
\end{equation}
The lightest mass eigenvalue can be calculated through the ratio of the determinant and the trace of \eqref{eq:Mass-matrix-scalar-real-block-2-times-2}, i.e.

\begin{equation}
\begin{split}
\frac{\mathrm{Det}\left[ M^{2}_{hH} \right]}{\mathrm{Tr}\left[ M^{2}_{hH} \right]} = \frac{m_{h}^{2}m_{H}^{2}}{m_{h}^{2}+m_{H}^{2}}\approx m_{h}^{2}
\end{split},
\end{equation}
obtaining 

\begin{equation}
m_{h}^{2} \approx \left( \tilde{\lambda}_{1}s_{\beta}^{2}+2\tilde{\lambda}_{5}c_{\beta}s_{\beta}
 + \tilde{\lambda}_{2}c_{\beta}^{2} \right){v^{2}},
\end{equation}
which we associate to the observed 125 GeV Higgs boson. The mixing angle associated to \eqref{eq:Mass-matrix-scalar-real-block-2-times-2} is defined as $t_{2\alpha}=\tan 2\alpha$, where

\begin{equation}
\label{eq:Mass-matrix-scalar-real-block-2-times-2-tangent}
t_{2\alpha} = \frac{f v_{\chi}+2\tilde{\lambda}_{5}s_{\beta}c_{\beta}v^{2} }
{f v_{\chi}+2 t_{2\beta}(s_{\beta}^{2}\tilde{\lambda}_{1}-c_{\beta}^{2}\tilde{\lambda}_{2})v_{2} }t_{2\beta}.
\end{equation}

Finally, the diagonalization of the CP-even matrix \eqref{eq:Mass-matrix-scalar-real} is achieved by $\mathit{R}_{\mathrm{R}}$ parametrized by a CKM-like matrix
\begin{equation}
\begin{split}
&\mathit{R}_{\mathrm{R}} = 
\begin{pmatrix}
1	&	0	&	0	\\	0	&	c_{23}	&	s_{23}	\\	0	&	-s_{23}	&	c_{23}
\end{pmatrix}
\begin{pmatrix}
c_{13}	&	0	&	s_{13}	\\	0	&	1	&	0	\\	-s_{13}	&	0	&	c_{13}
\end{pmatrix}
\begin{pmatrix}
c_{\alpha}	&	s_{\alpha}	&	0	\\	-s_{\alpha}	&	c_{\alpha}	&	0	\\	0	&	0	&	1
\end{pmatrix}
\end{split},
\end{equation}
where $t_{\alpha}=s_{\alpha}/c_{\alpha}$ and 

\begin{equation}
\begin{split}
s_{13} = \dfrac{1}{2}\frac{\lambda_{6} v s_{\beta} }{\lambda_{3}v_{\chi}} \, ,\quad 
s_{23} = \dfrac{1}{2}\frac{\lambda_{7} v c_{\beta} }{\lambda_{3}v_{\chi}} 	
\end{split},
\end{equation}
whose corresponding cosines are approximated as $c_{13}\approx 1-{s_{13}^{2}}/{2}$ and $c_{23}\approx 1-{s_{23}^{2}}/{2}$. In fact, the $R_{\mathrm{S}}$ matrix which block-diagonalizes $M_{\mathrm{R}}^{2}$ is the product of the former two rotation matrices with mixing angles $\theta_{23}$ and $\theta_{13}$. 

In conclusion, the scalar spectrum of the model is:
\begin{itemize}
\item Four would-be Goldstone bosons: $G_{W}^{\pm}$, $G_{Z}^{0}$ y $G_{Z'}^{0}$. 
\item Three scalar CP-even $h$, $H$ y $H_{\chi}$ fields with mass 
\begin{equation}
\begin{split}
m_{h}^{2} &\approx \left( \tilde{\lambda}_{1}c_{\beta}^{4}+2\tilde{\lambda}_{5}c_{\beta}^{2}s_{\beta}^{2}+\tilde{\lambda}_{2}s_{\beta}^{4} \right){v^{2}},	\\
m_{H}^{2} &\approx -\frac{f v_{\chi}}{4s_{\beta}c_{\beta}},	\\
m_{H_{\chi}}^{2} &\approx \lambda_{3} {v_{\chi}^{2}}.
\end{split}
\end{equation}
\item A pseudoscalar CP-odd $A^{0}$ whose mass is
\begin{equation}
m_{A^{0}}^{2}	= -\frac{1}{4}\frac{f v_{\chi}}{s_{\beta}c_{\beta}s_{\gamma}^{2}}.
\end{equation}
\item Two charged scalar bosons $H^{\pm}$ with mass
\begin{equation}
m_{H^{\pm}}^{2} = -\frac{1}{4}\frac{f v_{\chi}}{s_{\beta}c_{\beta}} -\frac{1}{4}\lambda_{5}' v^2.
\end{equation}
\end{itemize}

\subsection{Gauge boson masses}
\label{subsect:Vector-masses}
The kinetic terms of the scalar fields are
\begin{eqnarray}
\mathcal{L}_{kin} &=&\sum_i (D_{\mu}S)^{\dagger}(D^{\mu}S).
\label{higgs-kinetic}
\end{eqnarray}
After the symmetry breaking, the charged bosons $W_{\mu }^{\pm}=(W_{\mu }^{1}\mp W_{\mu }^{2})/\sqrt{2}$ acquire masses $M_{W}=g v /2$, while the masses for neutral gauge bosons are obtained from the following squared mass matrix in the basis $({W_{\mu }^3,B_{\mu},Z'_{\mu}})$:
\begin{equation}
M_0^2=\frac{1}{4}\begin{pmatrix}
g^2 v^2 &-gg' v^2   &-\frac{2}{3}gg_Xv^2(1+c_{\beta} ^2)   \\ 
&&\\
* &  g'^2v^2 &  \frac{2}{3}g'g_X v^2(1+c_{\beta}^2)   \\
 &&\\
* &*  &  \frac{4}{9}g_X^2 v_{\chi}^2\left[1+(1+3 c_{\beta}^2)\epsilon^2\right] \\
\end{pmatrix},
\end{equation}
where $\epsilon =\upsilon /\upsilon _{\chi }$. Taking into account $\epsilon ^2 \ll 1$, the matrix can be diagonalized with only two angles, obtaining the following mass eigenstates:
\begin{equation}
\begin{pmatrix}
A_{\mu} \\
Z_{1\mu } \\
Z_{2\mu } 
\end{pmatrix} \approx R_0
\begin{pmatrix}
W_{\mu}^3 \\
B_{\mu } \\
Z'_{\mu } 
\end{pmatrix},
\label{gauge-eigenvec}
\end{equation}  
with:
\begin{equation}
R_0 = \begin{pmatrix}
s_{W} & c_{W}  & 0 \\
c_{W} c_{Z} &  -s_{W} c_{Z}  &  s_{Z}   \\
-c_{W} s_{Z}  & s_W s_{Z}   & c_{Z}  \\
\end{pmatrix},
\end{equation}
where $\tan{\theta_{W}}=s_{W}/c_{W}=g'/g$ defines the Weinberg angle, and $s_{Z}=\sin{\theta_{Z}}$ is a small mixing angle between the SM neutral gauge boson $Z$ and the $\mathrm{U}(1)_X$ gauge boson $Z'$ such that in the limit $s_{Z}\rightarrow 0$, $Z_1=Z$ and $Z_2=Z'$. This mixing angle is approximately
\begin{equation}
s_{Z} \approx  (1+c_{\beta}^2)\frac{2g_X c_W}{3g}\left(\frac{M_Z}{M_{Z'}}\right)^2,
\label{mixing-angle}
\end{equation}
where the neutral masses are:
\begin{eqnarray}
M_Z\approx \frac{g\upsilon }{2c_W}\;, \ \ \  \  \ M_{Z'}\approx  \frac{g_X\upsilon _{\chi }}{3}.
\end{eqnarray}

\section{Fermion masses}
\label{sect:Fermion-masses}

\subsection{Quark sector}
\label{subsect:Quark-masses}
We find the Yukawa Lagrangian compatible with the $\mathrm{SU}(2)_L \otimes \mathrm{U}(1)_Y \otimes \mathrm{U}(1)_X$ gauge symmetry. For the quark sector we obtain

\begin{eqnarray}
-\mathcal{L}_Q &=& \overline{q_L^{1}}\left(\widetilde{\phi} _2h^{U}_2 \right)_{1j}U_R^{j}+\overline{q_L^{a}}(\widetilde{\phi }_1 h^{U}_{1})_{aj}U_R^{j}+\overline{q_L^{1}}\left(\phi _1 h^{D}_1\right)_{1j}D_R^{j}+\overline{q_L^{a}}\left(\phi _2 h^{D}_{2} \right)_{aj}D_R^{j} \notag \\
&+&\overline{q_L^{1}} (\phi _1 h^{J}_{1})_{1m} J^{m}_R+\overline{q_L^{a}}\left(\phi  _2 h^{J}_{2} \right)_{am} J^{m}_R+\overline{q_L^{1}}\left(\widetilde{\phi} _2 h^{T}_{2} \right)_1T_R +\overline{q_L^{a}} (\widetilde{\phi } _1 h^{T}_{1})_aT_R \notag \\
&+&\overline{T_{L}}\left( \sigma h_{\sigma }^{U}+\chi h_{\chi }^{U}\right)_{j}{U}_{R}^{j}+\overline{T_{L}}\left( \sigma h_{\sigma}^{T}+\chi h_{\chi }^{T}\right){T}_{R}
\nonumber \\
&+&\overline{J_{L}^n}\left( \sigma ^*h_{\sigma }^{D}+\chi ^*h_{\chi }^{D}\right)_{nj}{D}_{R}^{j}+\overline{J_{L}^n}\left( \sigma ^*h_{\sigma }^{J}+\chi ^*h_{\chi }^{J}\right)_{nm}{J}_{R}^{m}+h.c.,
 \label{quark-yukawa-1}
\end{eqnarray}
where $\widetilde{\phi}_{1,2}=i\sigma_2 \phi_{1,2}^*$ are conjugate fields,  $a=2,3$ label the second and third quark doublets and $n(m)=1,2$ is the index of the exotic $J^{n(m)}$ quarks.  A sum over the indices $i, a$ and $n$ is understood. 
 We can see in the quark Lagrangian that due to the non-universality of the $\mathrm{U}(1)_X$ symmetry, not all couplings between quarks and scalars are allowed by the gauge symmetry, which leads us to specific zero-texture Yukawa matrices. However, these structures are not inherited by the mass matrices of the quarks, due to the interactions of the four scalar fields $\phi _1$, $\phi _2$, $\sigma _0$ and $\chi _0$ that couple simultaneously to all quark flavors. In order to reproduce the observed mass spectrum, we must restrict further the number of couplings in the Lagrangian, which can be done by assuming the {\bf $Z_2$} discrete symmetries shown in tables \ref{tab:Fermionic-content} and \ref{tab:Scalar-content}.  Assuming these discrete symmetries, the Lagrangian (\ref{quark-yukawa-1}) after the symmetry breaking leads us to the following mass terms at tree level:
\begin{eqnarray}
-\langle\mathcal{L}_Q \rangle &=& \overline{U^{i}_{L}}(M_{U})_{ij}{U}_{R}^{j}+\overline{D^{i}_{L}}(M_{D})_{ij}{D}_{R}^{j}+\overline{T_{L}}(M_T){T}_{R}+\overline{J_{L}^n}(M_{J})_{nm}{J}_{R}^{m}
\nonumber \\
&+&\overline{T_{L}}(M_{TU})_{j}U_{R}^{j}+\overline{U^{i}_{L}}(M_{UT})_i{T}_{R} +\overline{D^{i}_{L}}(M_{DJ})_{im}J^m_R+ h.c.,
\label{mass-quarks-3}
\end{eqnarray}
where the mass matrices generate the following zero structures:
\begin{eqnarray}
M_{U}&=&\frac{1}{\sqrt{2}}\begin{pmatrix}
0 & \upsilon _2 a_{12} & 0 \\
0 & \upsilon _1 a_{22} & 0 \\
\upsilon _1 a_{31} & 0 & \upsilon _1a_{33} \\
\end{pmatrix}, \hspace{0.5cm}
M_{D}=\frac{ \upsilon _2}{\sqrt{2}}\begin{pmatrix}
0 & 0 & 0 \\
0 & 0 & 0 \\
B_{31} & B_{32} & B_{33} \\
\end{pmatrix},  \nonumber \\ \nonumber \\
M_J&=&\frac{ \upsilon _{\chi }}{\sqrt{2}}\begin{pmatrix}
k _{11} & k_{12}\\
k_{21} & k_{22} \\
\end{pmatrix}, \hspace{0.5cm}
M_T =\frac{\upsilon _{\chi}}{\sqrt{2}}h_{\chi }^T, \nonumber \\  \nonumber \\
M_{TU}&=&\frac{\upsilon _{\chi }}{\sqrt{2}}(0, c_2,0), \hspace{0.8cm} M_{UT}=\frac{1}{\sqrt{2}}\begin{pmatrix}
\upsilon _2 y_1 \\
\upsilon _1 y_2 \\
0 \\
\end{pmatrix} \nonumber \\
M_{DJ}&=&
\frac{1}{\sqrt{2}}\begin{pmatrix}
\upsilon _1 j_{11} & \upsilon _1 j_{12}\\
\upsilon _2 j_{21}   & \upsilon _2 j_{22} \\
0 & 0 \\
\end{pmatrix}, \hspace{0.8cm} M_{JD} = 0,
\label{mass-matrices-1}
\end{eqnarray}
which leads us to the following extended mass matrices:
\begin{eqnarray}
M'_{U}&=& \left( 
\begin{array}{ccc}
M_U  &\left| \right. &M_{UT} \\
\text{\textemdash \hspace{0.1cm} \textemdash} & \text{\textemdash}  &  \text{\textemdash  \hspace{0.1cm} \textemdash} \\ 
M_{TU}  &\left| \right. & M_T%
\end{array}%
\right)=\frac{1}{\sqrt{2}}\begin{pmatrix}
0 & \upsilon _2 a_{12} & 0 &  \left| \right. & \upsilon _2 y_1   \\
0 & \upsilon _1 a_{22} & 0 &  \left| \right. &  \upsilon _1 y_2 \\
\upsilon _1 a_{31} & 0 & \upsilon _1 a_{33} & \left| \right. &0\\
\text{\textemdash} & \text{\textemdash} & \text{\textemdash} & \text{\textemdash} & \text{\textemdash}\\
0  & \upsilon _{\chi}c_2  & 0   &  \left| \right.  & \upsilon _{\chi}h_{\chi }^T
\end{pmatrix}, \nonumber
\\  \nonumber 
\\
 M'_{D}&=&\left( 
\begin{array}{ccc}
M_D & \left| \right. &M_{DJ} \\
\text{\textemdash \hspace{0.1cm} \textemdash} & \text{\textemdash}  &  \text{\textemdash  \hspace{0.1cm} \textemdash} \\ 
M_{JD} & \left| \right. & M_J%
\end{array}%
\right)=\frac{1}{\sqrt{2}}\begin{pmatrix}
0 & 0 & 0 &  \left| \right.  & \upsilon _1 j_{11} & \upsilon _1 j_{12}   \\
0 & 0 & 0 &  \left| \right.  & \upsilon _2 j_{21} & \upsilon _2j_{22} \\
\upsilon _2 B_{31} & \upsilon _2 B_{32} & \upsilon _2 B_{33} & \left| \right.  & 0 &0 \\
 \text{\textemdash}  &  \text{\textemdash}  &  \text{\textemdash}  &  \text{\textemdash}  &  \text{\textemdash} &  \text{\textemdash}   \\
0  & 0  & 0  &  \left| \right.  & \upsilon _{\chi }k_{11} &  \upsilon _{\chi }k_{12} \\
0 & 0  &0 &  \left| \right. &  \upsilon _{\chi }k_{21}  &  \upsilon _{\chi }k_{22}
\end{pmatrix}.
\label{mass-matrices-2}
\end{eqnarray} 

After diagonalization, the above structures leads us to hierarchies of the phenomenological quarks, as detailed below.

\subsubsection*{Up sector}

First, we consider the up-type matrix $M'_U$ in equation (\ref{mass-matrices-2}). We obtain its symmetrical quadratic form as:

\begin{eqnarray}
\mathbb{M}_{U}^{2}&=&M'_U(M'_U)^{T} \nonumber \\
&=& \frac{1}{2}\begin{pmatrix}
\upsilon _2 ^2 \left(a_{12}^2+y_1^2 \right)& \upsilon _1 \upsilon _2 \left(a_{12}a_{22}+y_1y_2\right) & 0 &  \left| \right. & \upsilon _2 \upsilon _{\chi }\left(a_{12}c_2+y_1h_{\chi }^T\right)   \\
\upsilon _1 \upsilon _2 \left(a_{12}a_{22}+y_1y_2\right) &\upsilon _1 ^2 \left(a_{22}^2+y_2^2 \right) & 0 &  \left| \right. &  \upsilon _1 \upsilon _{\chi }\left(a_{22}c_2+y_2h_{\chi }^T\right) \\
0 & 0 & \upsilon _1 ^2 \left(a_{31}^2+a_{33}^2 \right) & \left| \right. &0\\
\text{\textemdash} & \text{\textemdash} & \text{\textemdash} & \text{\textemdash} & \text{\textemdash}\\
\upsilon _2 \upsilon _{\chi }\left(a_{12}c_2+y_1h_{\chi }^T\right)  & \upsilon _1 \upsilon _{\chi }\left(a_{22}c_2+y_2h_{\chi }^T\right)  & 0   &  \left| \right.  & \upsilon _{\chi}^2\left(c_2^2+h_{\chi }^{T2}\right)
\end{pmatrix}.\nonumber \\ 
\label{up-mass-tree}
\end{eqnarray}

The above mass matrix can be written as
\begin{eqnarray}
\mathbb{M}_{U}^{2}=\begin{pmatrix}
A & C \\
C^T & D
\label{up-mass-blocks}
\end{pmatrix},
\end{eqnarray}
which has the same structure as the general form of equation (\ref{block-matrix}) in the Appendix \ref{app:block}, where each block is:

\begin{eqnarray}
A &=& \frac{1}{2}\begin{pmatrix}
\upsilon _2 ^2 \left(a_{12}^2+y_1^2 \right) & \upsilon _1 \upsilon _2 \left(a_{12}a_{22}+y_1y_2\right)  &  0    \\
\upsilon _1 \upsilon _2 \left(a_{12}a_{22}+y_1y_2\right) &\upsilon _1 ^2 \left(a_{22}^2+y_2^2 \right) & 0  \\
0 & 0 &  \upsilon _1 ^2 \left(a_{31}^2+a_{33}^2 \right) 
\end{pmatrix}, \nonumber \\
   \nonumber \\
C &=&  \frac{1}{2}\begin{pmatrix}
 \upsilon _2 \upsilon _{\chi }\left(a_{12}c_2+y_1h_{\chi }^T\right)  \\
  \upsilon _1 \upsilon _{\chi }\left(a_{22}c_2+y_2h_{\chi }^T\right) \\
  0 
\end{pmatrix}, \nonumber \\ \nonumber \\
D &=& \frac{1}{2}\upsilon _{\chi}^2\left(c_2^2+h_{\chi }^{T2}\right).
\label{up-blocks}
\end{eqnarray}
We can see that each block are of the order $A \sim \upsilon _{1,2}^2$, $C \sim \upsilon _{1,2} \upsilon _{\chi}$ and $D \sim \upsilon _{\chi} ^2$, respectively, obeying the hierarchy from equation (\ref{block-hierarchy}). Thus, according to Appendix \ref{app:block}, the mass matrix (\ref{up-mass-blocks}) can be block diagonalized as

\begin{eqnarray}
\mathbbm{m} _U^2=\left(V_{L}^{(U)}\right)^T \mathbb{M}_{U}^{2} V_{L}^{(U)}=\begin{pmatrix}
m_U^2 & 0 \\
0 & m_T^2
\end{pmatrix},
\end{eqnarray}
where:

\begin{eqnarray}
m_U^2 &\approx & A-CD^{-1}C^T, \nonumber \\ 
m_T^2 &\approx & D,
\label{diagonal-up-blocks}
\end{eqnarray}

and the rotation matrix has the approximated form:

\begin{eqnarray}
V_{L}^{(U)}\approx \begin{pmatrix}
I & F_U \\
-F_U^T & I
\end{pmatrix}, \ \ \ \ \ \ F_U\approx CD^{-1}.
\end{eqnarray}

Since the block $D$ is just a number (see equation (\ref{up-blocks})), from (\ref{diagonal-up-blocks}) we obtain directly the mass of the heavy $T$ quark:
\begin{eqnarray}
m_T^2\approx  \frac{1}{2} \upsilon _{\chi}^2\left(c_2^2+h_{\chi }^{T2}\right).
\label{heavyT-mass}
\end{eqnarray}

On the other hand, from the matrices in (\ref{up-blocks}), and after some algebra, the matrix $m_U^2$ in (\ref{diagonal-up-blocks}), which contains the SM sector, can be put into the form:
\begin{eqnarray}
m_U^2 \approx  \frac{1}{2}\begin{pmatrix}
\upsilon _2^2  r_1^2 & \upsilon _1 \upsilon _2  r_1r_2 & 0 \\
\upsilon _1 \upsilon _2  r_1r_2 & \upsilon _1^2  r_2^2 & 0 \\
0 & 0 & \upsilon _1 ^2 \left(a_{31}^2+a_{33}^2 \right)
\end{pmatrix},
\label{SM-up-matrix}
\end{eqnarray}
where:
\begin{eqnarray}
r_1=	\frac{\left(a_{12}h_{\chi }^{T}-y_1c_2\right)}{\sqrt{c_2 ^2+h_{\chi }^{T2}}}, \nonumber \\
r_2=	\frac{\left(a_{22}h_{\chi }^{T}-y_2c_2\right)}{\sqrt{c_2 ^2+h_{\chi }^{T2}}}.
\label{ratio-couplings}
\end{eqnarray}
We see that the $33$ component of (\ref{SM-up-matrix}) appears decoupled, which corresponds to one of the eigenvalues. We associate this component to the top quark:

\begin{eqnarray}
m_t^2= \frac{1}{2}\upsilon _1 ^2 \left(a_{31}^2+a_{33}^2 \right),
\label{top-mass}
\end{eqnarray}
which leaves us with the $2\times 2$ submatrix

\begin{eqnarray}
m_{uc}^2 \approx  \frac{1}{2}\begin{pmatrix}
\upsilon _2^2  r_1^2 & \upsilon _1 \upsilon _2  r_1r_2  \\
\upsilon _1 \upsilon _2  r_1r_2 & \upsilon _1^2  r_2^2 
\end{pmatrix}.
\label{up-charm-matrix}
\end{eqnarray}
It is evident that the above matrix has null determinant, which leads us to at least one null eigenvalue. In fact, this structure produces one massless quark, which we associate to the lightest quark: the up quark ($u$), while the other eigenvalue, associated to the charm quark, corresponds to the trace of the matrix:
\begin{eqnarray}
m_c^2=\text{Tr}[m_{uc}^2]= \frac{1}{2}\left(\upsilon _1^2  r_2^2+\upsilon _2^2  r_1^2\right) \approx \frac{1}{2}\upsilon _1^2  r_2^2,
\label{charm-mass}
\end{eqnarray}
Since the mass of the top quark in (\ref{top-mass}) depends only on $\upsilon _1$, we take $\upsilon _2 \ll \upsilon _1$, which leads us to the approximation in equation (\ref{charm-mass}).

\begin{figure}[tb] 
\centering
\includegraphics[width=6cm,height=11cm,angle=-90]{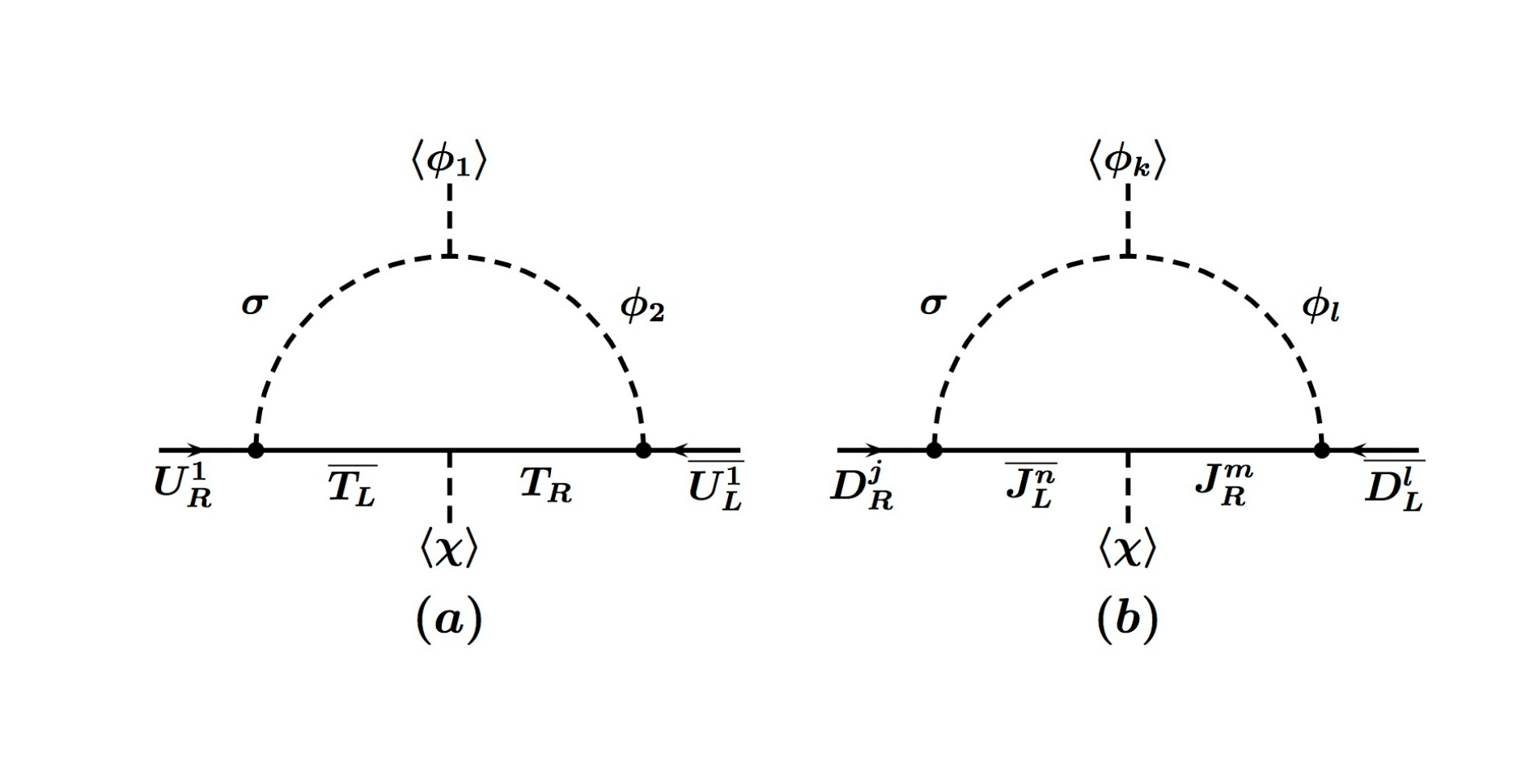}\vspace{-1cm}
\caption{\small Mass one-loop correction for ($a$) up and ($b$) down sector, where $k, l, m, n=1,2$ and $j=1,2,3$.}
\label{fig-oneloop}
\end{figure}

In order to generate mass to the $u$ quark, we consider the one-loop radiative correction shown in figure \ref{fig-oneloop}-(a). This contribution add an input into the $11$ component in the original $4\times 4$ matrix $M'_U$ in (\ref{mass-matrices-2}), which produces the one-loop quadratic mass matrix
\begin{eqnarray}
\mathbb{M}_{U(1)}^{2}=\mathbb{M}_{U}^{2}+\Delta \mathbb{M}_{U}^{2},
\label{up-mass-oneloop}
\end{eqnarray}
where the small one-loop contribution is:

\begin{eqnarray}
\Delta \mathbb{M}_{U}^{2}=
\frac{1}{2}\begin{pmatrix}
\upsilon _1 ^2 \Sigma _{11}^2 & 0 & \upsilon _1 ^2 a_{31} \Sigma _{11} &  \left| \right. & 0   \\
0 & 0 & 0 &  \left| \right. & 0 \\
\upsilon _1 ^2 a_{31} \Sigma _{11} & 0 & 0 & \left| \right. &0\\
\text{\textemdash} & \text{\textemdash} & \text{\textemdash} & \text{\textemdash} & \text{\textemdash}\\
0  & 0  & 0   &  \left| \right. & 0
\end{pmatrix},
\label{oneloop-shift}
\end{eqnarray}
and $\Sigma _{11}$ the value of the diagram in figure \ref{fig-oneloop}-(a) which obey the following analytical expression:

\begin{eqnarray}
\Sigma _{11}=\frac{-1}{16\pi ^2}\frac{f'\left(h^U_{\sigma}\right)_1\left(h^T_2\right)_1}{\sqrt{2}M_T}C_0\left(\frac{M_2}{M_T},\frac{M_{\sigma}}{M_T}\right),
\end{eqnarray} 
where:

\begin{eqnarray}
C_0\left(x_1,x_2\right)=\frac{1}{\left(1-x_1^2\right)\left(1-x_2^2\right)\left(x_1^2-x_2^2\right)}\left[x_1^2x_2^2\ln\left(\frac{{x_1^2}}{x_2^2}\right)-x_1^2\ln x_1^2+x_2^2 \ln x_2^2\right],
\label{oneloop-coef}
\end{eqnarray}
and $M_2$ is a charateristic mass arised from the internal $\phi _2$ line as linear combinations of mass eigevalues. The new one-loop contribution only has effect on the $3\times 3$ block matrix $m_U^2$ in (\ref{SM-up-matrix}), which change into the one-loop mass matrix
\begin{eqnarray}
m_{U(\text{1-loop})}^2 \approx  \frac{1}{2}\begin{pmatrix}
\upsilon _2^2  r_1^2+ \upsilon _1^2 \Sigma _{11}^2 & \upsilon _1 \upsilon _2  r_1r_2 & \upsilon _1^2 a_{31} \Sigma_{11}\\
\upsilon _1 \upsilon _2  r_1r_2 & \upsilon _1^2  r_2^2 & 0 \\
\upsilon _1^2 a_{31} \Sigma_{11} & 0 & 2m_t^2
\end{pmatrix},
\label{one-loop-SM-up-matrix}
\end{eqnarray}
where $m_t$ is the top mass at tree level obtained in (\ref{top-mass}). The new $13$ component emerged from the 1 loop diagram will correct the top mass. However, we will neglect this correction, which leads us again to a $2\times 2$ matrix
\begin{eqnarray}
m_{uc(\text{1-loop})}^2 \approx  \frac{1}{2}\begin{pmatrix}
\upsilon _2^2  r_1^2 + \upsilon _1^2 \Sigma _{11}^2 & \upsilon _1 \upsilon _2  r_1r_2  \\
\upsilon _1 \upsilon _2  r_1r_2 & \upsilon _1^2  r_2^2 
\end{pmatrix},
\label{up-charm-matrix-1loop}
\end{eqnarray}
which exhibits determinant different from zero. The trace of the matrix corresponds to the sum of the eigenvalues, i.e.:

\begin{eqnarray}
\text{Tr}[m_{uc(\text{1-loop})}^2]=m_u^2+m_c^2=\frac{1}{2}\left(\upsilon _1^2  r_2^2+\upsilon _2^2  r_1^2\right)+\frac{1}{2}\upsilon _1^2 \Sigma _{11}^2.
\end{eqnarray}

If we approximate the mass of the charm quark according to (\ref{charm-mass}), we obtain for the quark $u$ that:

\begin{eqnarray}
m_u^2=\frac{1}{2}\upsilon _1^2 \Sigma _{11}^2.
\label{up-mass}
\end{eqnarray}

\subsubsection*{Down sector}

For the down-type matrix $M'_D$ in (\ref{mass-matrices-2}), for simplicity we take in the heavy sector, proportional to $\upsilon _{\chi}$, a diagonal form, i.e., $k_{ij} =0$ for $i\neq j$. In this scenery, its quadratic form can also be put in the block form
\begin{eqnarray}
\mathbb{M}_{D}^{2}=\begin{pmatrix}
A & C \\
C^T & D
\label{down-mass-blocks}
\end{pmatrix},
\end{eqnarray}
where:
\begin{eqnarray}
A &=& \frac{1}{2}\begin{pmatrix}
\upsilon _1 ^2 \left(j_{11}^2+j_{12}^2 \right) & \upsilon _1 \upsilon _2 \left(j_{11}j_{21}+j_{12}j_{22}\right)  &  0    \\
\upsilon _1 \upsilon _2 \left(j_{11}j_{21}+j_{12}j_{22}\right) &\upsilon _2 ^2 \left(j_{21}^2+j_{22}^2 \right) & 0  \\
0 & 0 &  \upsilon _2 ^2 \left(B_{31}^2+B_{32}^2+B_{33}^2 \right) 
\end{pmatrix}, \nonumber \\
   \nonumber \\
C &=&  \frac{1}{2}\begin{pmatrix}
\upsilon _1 \upsilon _{\chi }j_{11}k_{11} &  \upsilon _1 \upsilon _{\chi }j_{12}k_{22}  \\
\upsilon _2 \upsilon _{\chi }j_{21}k_{11} &  \upsilon _2 \upsilon _{\chi }j_{22}k_{22} \\ 
0 & 0 
\end{pmatrix}, \nonumber \\ \nonumber \\
D &=& \frac{ \upsilon _{\chi }^2}{2}\begin{pmatrix}
k_{11}^2 & 0  \\
 0 &   k_{22}^2
\end{pmatrix}.
\label{down-blocks}
\end{eqnarray}
After block diagonalization, the matrix become:

\begin{eqnarray}
\mathbbm{m} _D^2=\left(V_{L}^{(D)}\right)^T \mathbb{M}_{D}^{2} V_{L}^{(D)}=\begin{pmatrix}
m_D^2 & 0 \\
0 & m_J^2
\end{pmatrix},
\end{eqnarray}
where:

\begin{eqnarray}
m_D^2 &\approx & A-CD^{-1}C^T, \nonumber \\ 
m_J^2 &\approx & D,
\label{diagonal-down-blocks}
\end{eqnarray}
with:

\begin{eqnarray}
V_{L}^{(D)}\approx \begin{pmatrix}
I & F_D \\
-F_D^T & I
\end{pmatrix}, \ \ \ \ \ \ F_D\approx CD^{-1}.
\end{eqnarray}
First, since the matrix $D$ appears diagonal, we obtain directly the mass of the heavy down-type quarks:

\begin{eqnarray}
m_{J^{1}}^2= \frac{1}{2}\upsilon _{\chi }^2k_{11}^2, \ \ \ \ m_{J^{2}}^2= \frac{1}{2}\upsilon _{\chi }^2k_{22}^2.
\end{eqnarray}
Second, for the SM down sector, the matrix $m_D^2$ in (\ref{diagonal-down-blocks}) gives:
\begin{eqnarray}
m_D^2=\frac{1}{2}\begin{pmatrix}
0 & 0 &  0    \\
0 & 0 & 0  \\
0 & 0 &  \upsilon _2 ^2 \left(B_{31}^2+B_{32}^2+B_{33}^2 \right) 
\end{pmatrix},
\label{SM-down-mass}
\end{eqnarray}
which exhibits two massless quarks: the down ($d$) and strange ($s$) quarks, and one massive quark associated to the bottom ($b$):
\begin{eqnarray}
m_b^2=\frac{1}{2}\upsilon _2 ^2 \left(B_{31}^2+B_{32}^2+B_{33}^2 \right).
\end{eqnarray}
In order to obtain mass for $d$ and $s$, we again consider the one-loop contribution shown in figure \ref{fig-oneloop}-(b), which produces new entrances different from zero in (\ref{SM-down-mass}) as follows:
\begin{eqnarray}
&&m_{D(\text{1-loop})}^2= \nonumber \\
&&\frac{1}{2}\begin{pmatrix}
\upsilon _2^2 \left(\Sigma _{11}^2+\Sigma _{12}^2+\Sigma _{13}^2\right) & \upsilon _1\upsilon _2 \left(\Sigma _{11}\Sigma _{21}+\Sigma _{12}\Sigma _{22}+\Sigma _{13}\Sigma _{23}\right) &  \upsilon _2^2 \left(\Sigma _{11}B_{31}+\Sigma _{12}B_{32}+\Sigma _{13}B_{33}\right)  \\
* & \upsilon _1^2 \left(\Sigma _{21}^2+\Sigma _{22}^2+\Sigma _{23}^2\right)  & \upsilon _1\upsilon _2 \left(\Sigma _{21}B_{31}+\Sigma _{22}B_{32}+\Sigma _{23}B_{33}\right)  \\
* & * &  2m_b^2
\end{pmatrix}, \nonumber \\
\label{oneloop-SM-down-mass}
\end{eqnarray}
where the one-loop correction is:

\begin{eqnarray}
\Sigma _{lj}=\frac{-1}{16\pi ^2}\frac{f'\left(h^J_l\right)_{lm}\left(h^D_{\sigma }\right)_{nj}}{\sqrt{2}M_J}C_0\left(\frac{M_l}{M_J},\frac{M_{\sigma}}{M_J}\right).
\end{eqnarray}

If the matrix in (\ref{oneloop-SM-down-mass}) is grouped as
\begin{eqnarray}
m_{D(\text{1-loop})}^2=\begin{pmatrix}
m_1^2 & n   \\ 
n^T &  2m_b^2
\end{pmatrix},\nonumber \\
\label{SM-down-mass-block}
\end{eqnarray}
where the bottom mass is dominant, we can block diagonalize it as:
\begin{eqnarray}
R_L^T m_{D(\text{1-loop})}^2R_L\approx \begin{pmatrix}
 m_{ds}^2&0 \\
 0&2m_b^2
\end{pmatrix},
\end{eqnarray}
with:
\begin{eqnarray}
m_{ds}^2&=&m_1^2-\frac{nn^T}{2m_b^2}  \nonumber \\
&=&\frac{1}{2m_b^2}\begin{pmatrix}
s_{11}\upsilon _2^2 & s_{12} \upsilon _1\upsilon _2 \\
s_{12} \upsilon _1\upsilon _2 & s_{22} \upsilon _1^2
\end{pmatrix},
\label{d-s-matrix}
\end{eqnarray}
and
\begin{eqnarray}
s_{11}&=&\left(\Sigma _{11}B_{32}-\Sigma _{12}B_{31}\right)^2+\left(\Sigma _{11}B_{33}-\Sigma _{13}B_{31}\right)^2+\left(\Sigma _{12}B_{33}-\Sigma _{13}B_{32}\right)^2, \nonumber \\
s_{22}&=&\left(\Sigma _{21}B_{32}-\Sigma _{22}B_{31}\right)^2+\left(\Sigma _{21}B_{33}-\Sigma _{23}B_{31}\right)^2+\left(\Sigma _{22}B_{33}-\Sigma _{23}B_{32}\right)^2, \nonumber \\
s_{12}&=&B_{31}^2\left(\Sigma _{13}\Sigma _{23}+\Sigma _{12}\Sigma _{32}\right)+B_{32}^2\left(\Sigma _{11}\Sigma _{12}+\Sigma _{13}\Sigma _{23}\right)+B_{33}^2\left(\Sigma _{11}\Sigma _{21}+\Sigma _{12}\Sigma _{22}\right) \nonumber \\
&-&B_{31}B_{32}\left(\Sigma _{12}\Sigma _{21}+\Sigma _{11}\Sigma _{22}\right)-B_{31}B_{33}\left(\Sigma _{11}\Sigma _{23}+\Sigma _{13}\Sigma _{21}\right)-B_{32}B_{33}\left(\Sigma _{13}\Sigma _{22}+\Sigma _{12}\Sigma _{23}\right). \nonumber \\
\end{eqnarray}
The eigenvalues of $m_{ds}^2$ in (\ref{d-s-matrix}) will lead us to the down and strange masses. For example, if the mixing component $s_{12}$ is null, we obtain:
\begin{eqnarray}
m_d^2\approx \frac{s_{11}\upsilon _2^2}{2m_b^2}, \nonumber \\
m_s^2\approx \frac{s_{22}\upsilon _1^2}{2m_b^2}.
\end{eqnarray}

\subsection{Lepton sector}
\label{subsect:Lepton-masses}
The non-universal $U(1)_{X}$ also forbids some Yukawa couplings between leptons and scalar bosons. The allowed couplings are shown below for neutral and charged leptons, respectively:

\begin{equation}
\begin{split}
-\mathcal{L}_{Y,N} &= 
h_{2e}^{\nu e}\overline{\ell^{e}_{L}}\tilde{\phi}_{2}\nu^{e}_{R} + 
h_{2e}^{\nu \mu}\overline{\ell^{e}_{L}}\tilde{\phi}_{2}\nu^{\mu}_{R} + 
h_{2e}^{\nu \tau}\overline{\ell^{e}_{L}}\tilde{\phi}_{2}\nu^{\tau}_{R} + 
h_{2\mu}^{\nu e}\overline{\ell^{\mu}_{L}}\tilde{\phi}_{2}\nu^{e}_{R} +
h_{2\mu}^{\nu \mu}\overline{\ell^{\mu}_{L}}\tilde{\phi}_{2}\nu^{\mu}_{R} + 
h_{2\mu}^{\nu \tau}\overline{\ell^{\mu}_{L}}\tilde{\phi}_{2}\nu^{\tau}_{R} \\ &+
h_{\chi i}^{\nu j} \overline{\nu_{R}^{i\;C}} \chi^{*} N_{R} +
\frac{1}{2} \overline{N_{R}^{i\;C}} M^{ij}_{N} N_{R}^{j} + \mathrm{h.c.},
\end{split}
\label{eq:Neutrino-Lagrangian}
\end{equation}

\begin{equation}
\begin{split}
-\mathcal{L}_{Y,E} &= 
\eta \overline{\ell^{e}_{L}}\phi_{2}e^{\mu}_{R} + h \overline{\ell^{\mu}_{L}}\phi_{2}e^{\mu}_{R} + 
\zeta\overline{\ell^{\tau}_{L}}\phi_{2}e^{e}_{R} + H\overline{\ell^{\tau}_{L}}\phi_{2}e^{\tau}_{R} +	
q_{11}\overline{\ell^{e}_{L}}\phi_{1}E_{R} + q_{21}\overline{\ell^{\mu}_{L}}\phi_{1}{E}_{R} \\ &+
h_{\sigma e}^{E}\overline{E_{L}}\sigma e^{e}_{R} + h_{\sigma \mu}^{\mathcal{E}}\overline{\mathcal{E}_{L}}\sigma^{*} e^{\mu}_{R} + 
h_{\sigma \tau}^{E}\overline{E_{L}}\sigma e^{\tau}_{R} + 
H_{1}\overline{E_{L}}\chi E_{R} + H_{2}\overline{\mathcal{E}_{L}}\chi^{*} \mathcal{E}_{R} + \mathrm{h.c.}
\end{split}
\label{eq:Electron-Lagrangian}
\end{equation}

Since the Higgs doublet $\phi_{2}$ has the discrete symmetry $\phi_{2}\rightarrow -\phi_{2}$, all the right-handed leptons except $E_{R}$ and $\mathcal{E}_{R}$ also have $\mathbf{Z}_{2}$ negative parities in order to obtain the adequate zero textures, i.e.:
\begin{equation}
 e ^{e,\mu,\tau}_{R}\rightarrow - e ^{e,\mu,\tau}_{R}, \qquad
\nu^{e,\mu,\tau}_{R}\rightarrow -\nu^{e,\mu,\tau}_{R}, \qquad
 N ^{e,\mu,\tau}_{R}\rightarrow - N ^{e,\mu,\tau}_{R}.
\end{equation}

\subsubsection*{Neutral leptons}
Evaluating in the VEVs, the terms obtained from \eqref{eq:Neutrino-Lagrangian} can be written in the following mass term using the basis $\mathbf{N}_{L}=\left(\begin{matrix}{\nu^{e,\mu,\tau}_{L}},\,\left(\nu^{e,\mu,\tau}_{R}\right)^{C},\,\left(N^{e,\mu,\tau}_{R}\right)^{C}\end{matrix}\right)^{\mathrm{T}}$ for the neutral sector

\begin{equation}
-\mathcal{L}_{Y,N} = \frac{1}{2} \overline{\mathbf{N}_{L}^{C}} \mathbb{M}_{\nu} \mathbf{N}_{L},
\end{equation}
where the mass matrix is

\begin{equation}
\mathbb{M}_{\nu} = 
\left(\begin{array}{c c c}
0	&	m_{D}^{\mathrm{T}}	&	0	\\
m_{D}	&	0	&	M_{D}^{\mathrm{T}}	\\
0	&	M_{D}	&	M_{M}
\end{array}\right),
\end{equation}
with $M_{D}=h_{\chi}^{\nu}v_{\chi}/\sqrt{2}$ being a Dirac mass between $\nu_{R}^{c}$ and $N_{R}$, where $h_{N\chi}$ is a $3\times 3$ matrix, and

\begin{equation}
\label{eq:m_nu_original_parameters}
m_{D} = \frac{v_{2}}{\sqrt{2}}\left(\begin{matrix}
h_{2e}^{\nu e}	&	h_{2e}^{\nu \mu}	&	h_{2e}^{\nu \tau}	\\
h_{2\mu}^{\nu e}&	h_{2\mu}^{\nu \mu}	&	h_{2\mu}^{\nu \tau}	\\
0	&	0	&	0	\end{matrix}\right),
\end{equation}
is a Dirac mass matrix between $\nu_{L}$ and $\nu_{R}$. $M_{M}$ is the mass of the Majorana neutrino $N_{R}$.

Considering that $M_M \ll m_D$ and $M_D $, the matrix $\mathbb{M}_{\nu}$ can be diagonalized through the inverse seesaw mechanism \cite{inverseseesaw,catano2012}. If the following blocks are defined
\begin{equation}
\begin{split}
\mathcal{M}_{\nu} &= \left(\begin{matrix}
m_{D}	\\	0
\end{matrix} \right),\\
\mathcal{M}_{N} &= \left(\begin{matrix}
0	&	M_{D}^{\mathrm{T}}	\\
M_{D}	&	M_{M}
\end{matrix} \right),
\end{split}
\end{equation}
the mass matrix becomes

\begin{equation}
\mathbb{M}_{\nu} = \left(\begin{matrix}
0	&	\mathcal{M}_{\nu}^{\mathrm{T}}	\\
\mathcal{M}_{\nu}	&	\mathcal{M}_{N}
\end{matrix} \right),
\end{equation}
which has the same form as the block matrix (\ref{block-matrix}) from Appendix \ref{app:block} in the limit with $A=0$. Thus, we define the rotations

\begin{equation}
{\mathbb{W}_{\mathrm{SS}}}^{\mathrm{T}} \mathbb{M}_{\nu} \mathbb{W}_{\mathrm{SS}} = \left( \begin{matrix}
m_{\mathrm{light}}	&	0	\\	0	&	m_{\mathrm{heavy}}
\end{matrix} \right),
\end{equation}
with
 
\begin{equation}
\mathbb{W}_{\mathrm{SS}} \approx \left( \begin{matrix}
I	&	F^{N}	\\
-\left(F^{N}\right)^T	&	I
\end{matrix} \right), \ \ \ \ F^{N} \approx \left( \mathcal{M}_{N} \right)^{-1} \mathcal{M}_{\nu},
\label{seesaw-rotation}
\end{equation}
and

\begin{eqnarray}
m_{\mathrm{light}} &\approx &	- \mathcal{M}_{\nu}^{\mathrm{T}} \mathcal{M}_{N}^{-1} \mathcal{M}_{\nu}	,\\
m_{\mathrm{heavy}} &\approx &	\mathcal{M}_{N}.
\end{eqnarray}

Since

\begin{equation}
\mathcal{M}_{N}^{-1} = \left(\begin{matrix}
-\left( M_{D} \right)^{-1} M_{M} \left( M_{D}^{\mathrm{T}} \right)^{-1}	&	M_{D}^{-1}	\\
\left( M_{D}^{\mathrm{T}} \right)^{-1} &	0
\end{matrix} \right),
\end{equation}
the light mass term is
\begin{equation}
\label{eq:Light-neutrino-mass-matrix}
m_{\mathrm{light}} =	m_{D}^{\mathrm{T}} \left( M_{D} \right)^{-1} M_{M} \left( M_{D}^{\mathrm{T}} \right)^{-1} m_{D}.
\end{equation}

Now, a unitary matrix $\mathbb{V}$ is considered which diagonalizes the $3 \times 3$ block matrix $\mathcal{M}_{N}$ \cite{catano2012}:

\begin{equation}
\begin{split}
\mathbb{V}^{\mathrm{T}}\mathcal{M}_{N}\mathbb{V} &= \mathbb{V}^{\mathrm{T}} \left( \begin{matrix}
0	&	M_{D}	\\
M_{D}^{\mathrm{T}} &	M_{N}
\end{matrix} \right) \mathbb{V} \\
 &= \left( \begin{matrix}
V_{1}^{*}M^{\mathrm{diag}}_{1}V_{1}^{\dagger}	&	0	\\
0	&	V_{2}^{*}M^{\mathrm{diag}}_{2}V_{2}^{\dagger}
\end{matrix} \right),
\end{split}
\label{block-diagonalization}
\end{equation}
with $V_{1}$ and $V_{2}$  sub-rotation matrices. $\mathbb{V}$ may be formally expressed as \cite{catano2012}

\begin{equation}
\mathbb{V} = \frac{1}{\sqrt{2}}\left( \begin{matrix}
1	&	1	\\	-1	&	1
\end{matrix} \right) \left( \begin{matrix}
1 - \frac{SS^{\dagger}}{2}	&	S	\\	-S^{\dagger}	&	1 - \frac{S^{\dagger}S}{2}
\end{matrix} \right).
\end{equation}
Using (\ref{block-diagonalization}), and assuming that $M_{D}=M_{D}^{\mathrm{T}}$, $M_{M}S^{\dagger}=S^{\mathrm{T}}M_{M}$, $M_{M}S=S^{*}M_{M}$, $M_{D}S^{\dagger}=S^{\mathrm{T}}M_{D}$ and $M_{D}S=S^{*}M_{D}$, from the off-diagonal elements we find
\begin{equation}
\begin{split}
S = S^{\dagger} =& -\frac{1}{4}M_{D}^{-1}M_{M},
\end{split}
\end{equation}
and substituting for the diagonal elements, we get the mass matrices
\begin{eqnarray}
V_{1}^{*}M^{\mathrm{diag}}_{1}V_{1}^{\dagger} &=& \frac{M_{M}}{2} - M_{D} - \frac{1}{8}M_{M}M_{D}^{-1}M_{M}\approx -M_{D} ,	\\
V_{2}^{*}M^{\mathrm{diag}}_{2}V_{2}^{\dagger} &=& \frac{M_{M}}{2} + M_{D} +\frac{1}{8}M_{M}M_{D}^{-1}M_{M} \approx M_{D}.
\end{eqnarray}

The mass eigenstates $\mathbf{n}_{L}$ are constructed as:
\begin{equation}
\label{eq:Total-neutrino-mixing}
\mathbf{N}_{L} = \mathbb{U}_{N}\mathbf{n}_{L},
\end{equation}
with $\mathbf{n}_{L}=\left(\nu^{1,2,3}_{L},\, N^{1,2,3}_{1L},\, N^{1,2,3}_{2L} \right)$, and the rotation matrix as
\begin{equation}
\label{eq:Total-neutrino-mixing-matrix}
\mathbb{U}_{N} = \mathbb{W}_{\mathrm{SS}}\mathbb{W}_\mathrm{H}\mathbb{W}_{\mathrm{B}},
\end{equation}
with $\mathbb{W}_{\mathrm{SS}}$ the seesaw matrix rotation from (\ref{seesaw-rotation}),
\begin{equation}
\mathbb{W}_{\mathrm{H}} = \left( \begin{matrix}
1	&	0	\\
0	&	\mathbb{V}
\end{matrix} \right)
\end{equation}
the matrix rotation of the heavy neutrinos, and
\begin{equation}
\mathbb{W}_{\mathrm{B}} = \mathrm{block\,diag} \left( U_{\nu},\, V_{1},\, V_{2} \right)
\end{equation}
the matrices that diagonalize each $3 \times 3$ block.

\subsubsection*{Charged leptons}
\begin{figure}[tb] 
\centering
\includegraphics[width=5cm,height=9cm,angle=-90]{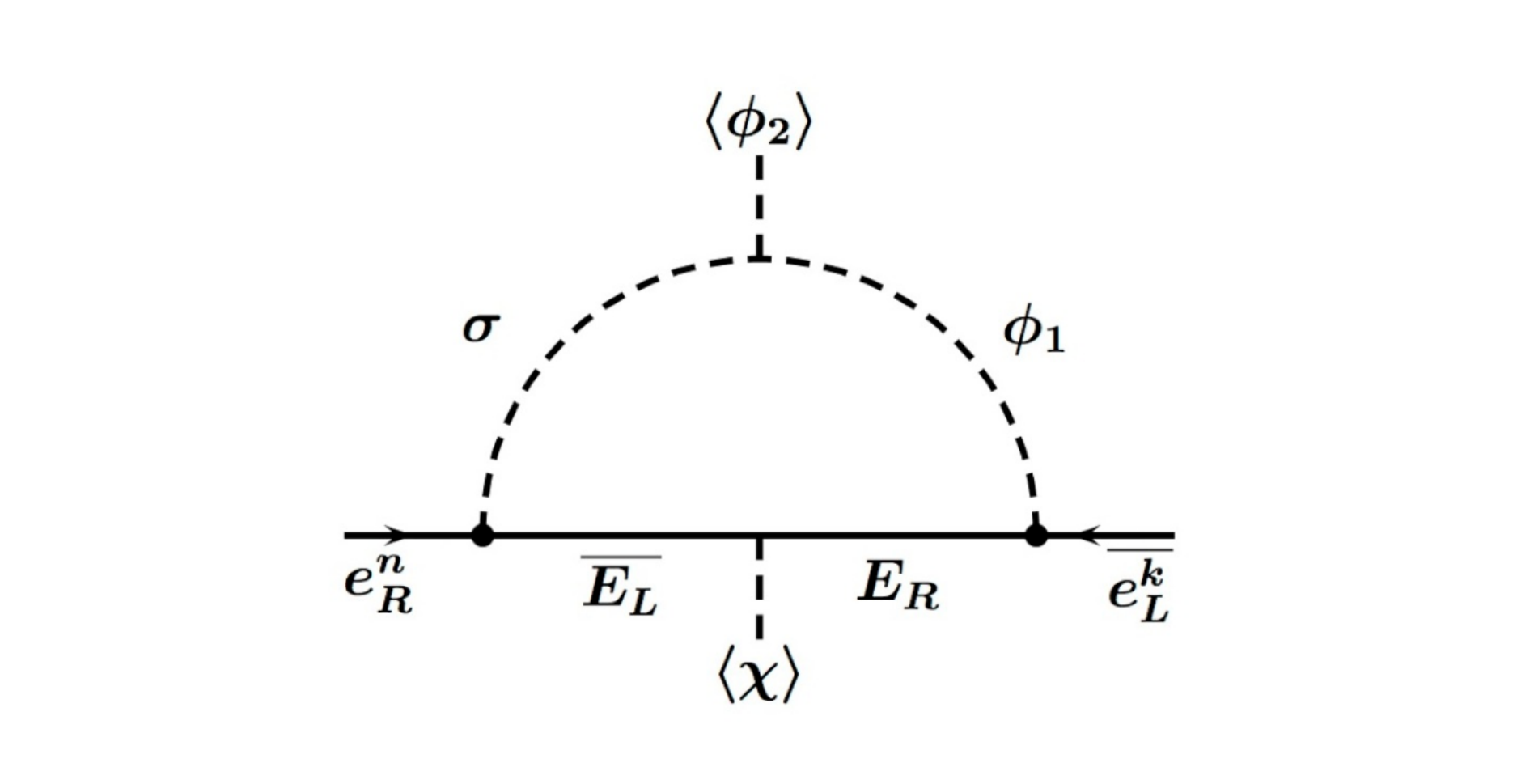}\vspace{-0.5cm}
\caption{\small Mass one-loop correction for charged leptons, where $n=e,\tau $ and $k=e,\mu$.}
\label{fig-oneloop-2}
\end{figure}

For the charged sector in the flavor basis $\mathbf{E}=(e^{e},e^{\mu} , e^{\tau}, E)$, the mass terms obtained from \eqref{eq:Electron-Lagrangian} after the symmetry breaking are

\begin{equation}
-\mathcal{L}_{Y,E} = \overline{\mathbf{E}_{L}}\mathbb{M}_{E} \mathbf{E}_{R}
 + \frac{H_{2}v_{\chi}}{\sqrt{2}}\overline{\mathcal{E}_{L}}\mathcal{E}_{R} + \mathrm{h.c.},
\end{equation}
where the lepton mass matrix $\mathbb{M}_{E}$ has de following form:

\begin{equation}
\mathbb{M}_{E} = \frac{v_{2}}{\sqrt{2}}
\begin{pmatrix}
 0  & \eta & 0 & | & q_{11} t_{\beta} \\
 0 & h  & 0 & | & q_{21} t_{\beta} \\
 \zeta        & 0 & H & | & 0 \\
 - & - & - & -& - \\
 0 & 0 & 0 & | &H_{1} v_{\chi}/v_{2}
\end{pmatrix},
\end{equation}
which exhibits one massless lepton (the electron). To obtain a massive electron, we include the one-loop correction shown in figure \ref{fig-oneloop-2}, which add a new term

\begin{eqnarray}
\mathbb{M}_{E(\text{1})}=\mathbb{M}_{E}+\Delta \mathbb{M}_{E},
\end{eqnarray}
with:

\begin{eqnarray}
\Delta \mathbb{M}_{E}=
\frac{\upsilon _2 }{2}\begin{pmatrix}
\Sigma _{11} & 0 &  \Sigma _{13} &  \left| \right. & 0   \\
\Sigma _{21} & 0 & \Sigma _{23} &  \left| \right. & 0 \\
0 & 0 & 0 & \left| \right. &0\\
\text{\textemdash} & \text{\textemdash} & \text{\textemdash} & \text{\textemdash} & \text{\textemdash}\\
0  & 0  & 0   &  \left| \right. & 0
\end{pmatrix}.
\label{oneloop-shift-lept}
\end{eqnarray}

Since $\mathbb{M}_{E(1)}$ is not hermitian, there are two rotation matrices $\mathbb{V}^{E}_{L}$ and $\mathbb{V}^{E}_{R}$ for left- and right-handed electrons. Hence, the left-handed rotation is obtained by diagonalizing $\mathbb{M}_{E}\mathbb{M}_{E}^{\dagger}$
obtaining the corresponding eigenvalues
\begin{equation}
\begin{split}
m_{e}^{2} &= \frac{h^2 \Sigma_{11}^2 v_{2}^2}{2 \left(\eta ^2+h^2\right)}\approx \frac{ v_{2}^2}{2}\Sigma_{11}^2,	\\
m_{\mu}^{2} &= \frac{v_{2}^{2}}{2} \left(\eta^2 + h^2\right)\approx \frac{v_{2}^{2}}{2} h^2,	\\
m_{\tau}^{2} &= \frac{v_{2}^{2}}{2} \left(\zeta^2 + H^2\right)\approx \frac{v_{2}^2}{2}H^2,	\\
m_{E}^{2} &= \frac{H_{1}^2 v_{\chi}^2}{2}.
\end{split}
\end{equation}

In addition, the flavor eigenstates are related to mass eigenstates $\mathbf{e}=(e, \mu, \tau, E')^{\mathrm{T}}$ by:

\begin{equation}
\mathbf{E}_{L} = \mathbb{V}^{E}_{L}\mathbf{e}_{L}
\end{equation}
where the corresponding left-handed rotation matrix can be expressed as:

\begin{equation}
\mathbb{V}^{E}_{L} = \mathbb{V}_{\mathrm{SS},L}^{E}\mathbb{V}_{\mathrm{SM},L}^{E},
\end{equation}
which diagonalizes as:

\begin{equation}
\mathbb{M}_{E}\mathbb{M}_{E}^{\dagger} = \frac{1}{2}
\begin{pmatrix}
\mathcal{M}_{ee}^{2}	&	\mathcal{M}_{eE}^{2}	\\
{\mathcal{M}_{eE}^{2\;\mathrm{T}}}	&	\mathcal{M}_{EE}^{2}
\end{pmatrix},
\end{equation}
whose blocks are

\begin{equation}
\begin{split}
\mathcal{M}_{ee}^{2} &= \frac{v_{2}^{2}}{2}
\begin{pmatrix}
	{q_{11}^2 t_{\beta}^2}+{\eta ^2}+{  \Sigma_{11}^2}+{  \Sigma_{13}^2} & 
	q_{11} q_{21} t_{\beta}^{2}+h   \eta +   \Sigma_{11} \Sigma_{21}+  \Sigma_{13} \Sigma_{23} & 
	\zeta  \Sigma_{11}  +H \Sigma_{13}   \\
	* & {q_{21}^2 t_{\beta}^{2}}+{h^2  }+{  \Sigma_{21}^2}+{  \Sigma_{23}^2} & 
	\zeta  \Sigma_{21}  + H \Sigma_{23}   \\
	* & 
	* & H^2 + \zeta^2
\end{pmatrix},\\
\mathcal{M}_{eE}^{2} &= \frac{v_{1} v_{\chi}}{2} H_{1}
\begin{pmatrix}
 q_{11}	 \\
 q_{21}	 \\
 0	 
\end{pmatrix},\\
\mathcal{M}_{EE}^{2} &= \frac{v_{\chi}^{2}H_{1}}{2}.
\end{split}
\end{equation}
The former matrix $\mathbb{V}_{\mathrm{SS},L}^{E}$ is

\begin{equation}
\mathbb{V}_{\mathrm{SS},L}^{E} =
\begin{pmatrix}
I		&	F^{E}	\\
-F^{E\dagger}	&	I
\end{pmatrix},
\end{equation}
with $F^{E}=\mathcal{M}_{eE}^{2}\left(\mathcal{M}_{EE}^{2}\right)^{-1} $. The latter rotation is:

\begin{equation}
\mathbb{V}_{\mathrm{SM},L}^{E} = 
\begin{pmatrix}
{V}_{\mathrm{SM},L}^{E}	&	0	\\	0	&	1
\end{pmatrix},
\end{equation}
where the top-left block diagonalizes the SM charged lepton masses

\begin{equation}
\label{eq:electron-muon-rotation}\
{V}_{\mathrm{SM},L}^{E} = 
\begin{pmatrix}
 c_{\alpha_{e\mu}}	&	s_{\alpha_{e\mu}}	&	\frac{\Sigma_{13}}{H}	\\
-s_{\alpha_{e\mu}}	&	c_{\alpha_{e\mu}}	&	\frac{\Sigma_{23}}{H}	\\
-\frac{\Sigma_{13}}{H}	&	-\frac{\Sigma_{23}}{H}	&	1
\end{pmatrix}.
\end{equation}
The angle $\alpha_{e\mu}$ is defined by $t_{\alpha_{e\mu}} = \tan \alpha_{e\mu} \approx \eta/h$, which is a free parameter of the model as shown below.

\section{PMNS matrix}
\label{sect:PMNS-matrix}

To explore some phenomenological consequences of the above structures, we assume for simplicity that $M_{D}$ is diagonal and $M_{M}$ is proportional to the identity

\begin{equation}
M_{D} = \left( \begin{matrix}
h_{N\chi 1}	&	0	&	0	\\	0	&	h_{N\chi 2}	&	0	\\	0	&	0	&	h_{\chi N 3}
\end{matrix} \right)\frac{v_{\chi}}{\sqrt{2}}
\end{equation}
\begin{equation}
M_{M} = \mu_{N} \mathbb{I}_{3\times 3}.
\end{equation}
Thus, $V_{1}=V_{2}=\mathbb{I}_{3\times 3}$ in (\ref{block-diagonalization}). On the other hand, replacing the Dirac matrix from \eqref{eq:m_nu_original_parameters} into the light mass eigenvalues in \eqref{eq:Light-neutrino-mass-matrix}, we obtain

\begin{equation}
\label{eq:Neutrino-mass-matrix}
m_{\mathrm{light}} = \frac{\mu_{N} v_{2}^{2}}{{h_{N\chi 1}}^{2}v_{\chi}^{2}}
\left( \begin{matrix}
	\left( h_{2e}^{\nu e}\right)^{2} + \left( h_{2\mu}^{\nu e} \right)^{2} \rho^{2} 	&
	{h_{2e}^{\nu e}}\,{h_{2e}^{\nu \mu}} + {h_{2\mu}^{\nu e}}\,{h_{2\mu}^{\nu \mu}}\rho^2 	&
	{h_{2e}^{\nu e}}\,{h_{2e}^{\nu \tau}}+ {h_{2\mu}^{\nu e}}\,{h_{2\mu}^{\nu \tau}}\rho^2 	\\
	{h_{2e}^{\nu e}}\,{h_{2e}^{\nu \mu}} + {h_{2\mu}^{\nu e}}\,{h_{2\mu}^{\nu \mu}}\rho^2	&	
	\left( h_{2e}^{\nu \mu} \right)^{2} + \left( h_{2\mu}^{\nu\mu} \right)^{2} \rho^{2}	&	
	{h_{2e}^{\nu \mu}}\,{h_{2e}^{\nu \tau}}+ {h_{2\mu}^{\nu \mu}}\,{h_{2\mu}^{\nu \tau}}\rho^2	\\
	{h_{2e}^{\nu e}}  \,{h_{2e}^{\nu \tau}}+ {h_{2\mu}^{\nu e}}  \,{h_{2\mu}^{\nu \tau}}\rho^2	&	
	{h_{2e}^{\nu \mu}}\,{h_{2e}^{\nu \tau}}+ {h_{2\mu}^{\nu \mu}}\,{h_{2\mu}^{\nu \tau}}\rho^2	&	
	\left( h_{2e}^{\nu \tau} \right)^{2} + \left( h_{2\mu}^{\nu \tau} \right)^{2} \rho^{2}
\end{matrix} \right),
\end{equation}
where $\rho={h_{N\chi 1}}/{h_{N\chi 2}}$. The matrix $m_{\mathrm{light}}$ has zero determinant, obtaining at least, one massless neutrino. The above matrix is diagonalized through

\begin{equation}
U_{\mathrm{\nu}}^{\mathrm{T}}\, m_{\mathrm{light}}\, U_{\nu} = m_{\mathrm{light}}^{\mathrm{diag}},
\end{equation}
where $U_{\mathrm{\nu}}$ contains the mixing angles that transform the weak eigenstates $\nu_{L}^{e,\mu,\tau}$ into mass eigenstates $\nu_{L}^{1,2,3}$. The PMNS matrix is defined as the product of the above rotation matrix and the rotation matrix of the charged sector ${V}_{\mathrm{SM},L}^{E}$

\begin{equation}
U_{\mathrm{PMNS}} = \left( {V}_{\mathrm{SM},L}^{E} \right)^{\dagger} U_{\nu}.
\end{equation}

We use the following parametrization for the PMNS matrix \cite{Beringer2012}:

\begin{equation}
U_{\mathrm{PMNS}} = \left( \begin{matrix}
c_{12}c_{13}	&	s_{12}c_{13}	&	s_{13}e^{-i\delta}	\\
-s_{12}c_{23}-c_{12}s_{23}s_{13}e^{i\delta}	&	
  c_{12}c_{23}-s_{12}s_{23}s_{13}e^{i\delta}	&	s_{23}c_{13}	\\
 s_{12}s_{23}-c_{12}c_{23}s_{13}e^{i\delta}	&	
-c_{12}s_{23}-s_{12}c_{23}s_{13}e^{i\delta}	&	c_{23}c_{13}
\end{matrix} \right).
\end{equation}
The mixing angles can be obtained from some matrix componenets as

\begin{equation}
\begin{split}
s_{13}^{2} &= \left| U_{e3} \right|^{2},	\\
s_{23}^{2} &= \frac{\left| U_{\mu3} \right|^{2}}{1-\left| U_{e3} \right|^{2}},	\\
s_{12}^{2} &= \frac{\left| U_{  e2  } \right|^{2}}{1-\left| U_{e3} \right|^{2}}.
\end{split}
\label{PMNS-mixing angles}
\end{equation}

\subsubsection*{Parameter values }

In order to has a model consistent with neutrino oscillation data \cite{neutrinodata}, the values of the Yukawa parameters $h_{2e}^{\nu e}$, $h_{2e}^{\nu \mu}$, $h_{2e}^{\nu \tau}$, $h_{2\mu}^{\nu e}$, $h_{2\mu}^{\nu \mu}$, $h_{2\mu}^{\nu \tau}$ and $\alpha_{e\mu}$ must be properly adjusted. To achieve this, we implement a MonteCarlo method to generate random numbers in the parameter space, where only the numbers which match up the mass matrix to experimental data are accepted, while the others are rejected. It is worth mentioning that the other two rotation parameters described by $\Sigma_{13}/H$ and $\Sigma_{23}/H$ were approximated to $m_{e}/m_{\tau}$, while $h_{2e}^{\nu \mu}$ was chosen null to simplify the search.

On the other hand, the appropriate mass scale and mass ordering can be obtained by adjusting the outer factor of the mass matrix and the ratio $\rho$. For NO the Yukawa coupling can be set by
\begin{equation}
\begin{split}
{h_{N\chi 1}}^{2} &= 0.02,	\\
{\rho}^{2} &= 0.5,
\end{split}
\end{equation}
while for IO
\begin{equation}
\begin{split}
{h_{N\chi 1}}^{2} &= 0.025,	\\
{\rho}^{2} &= 0.625.	
\end{split}
\end{equation}
In the same way, the mass scale is set by
\begin{equation}
\begin{split}
v_{2} = 7\mathrm{\,GeV},	\\
v_{\chi} = 7\mathrm{\,TeV},	\\
\mu_{N} = 1\mathrm{\,keV}.
\end{split}
\end{equation}
The above values fix the outer factor of the mass matrix \eqref{eq:Neutrino-mass-matrix} at $50\mathrm{\,meV}$, which yields to the correct squared-mass differences. Nevertheless, there exist other possible values for the parameters $\mu_{N}$, $h_{N1\chi}$, $v_{\chi}$ and $\tan \beta$ that lead us the factor at $50$ meV. 

\begin{figure}[htbp]
\centering
\subfigure[NO: ${h_{N\chi 1}}^{2} = 0.01$.]{\includegraphics[width=40mm]{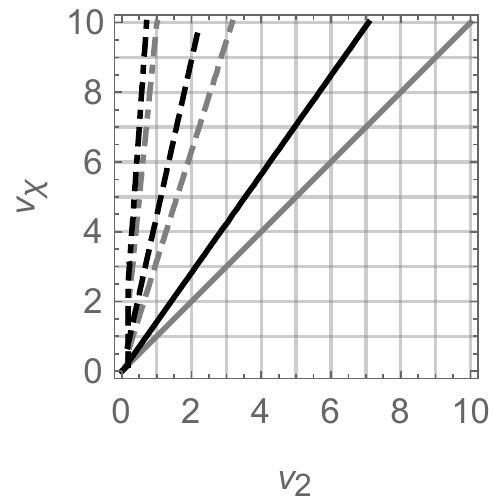}}
\subfigure[NO: ${h_{N\chi 1}}^{2} = 0.10$.]{\includegraphics[width=40mm]{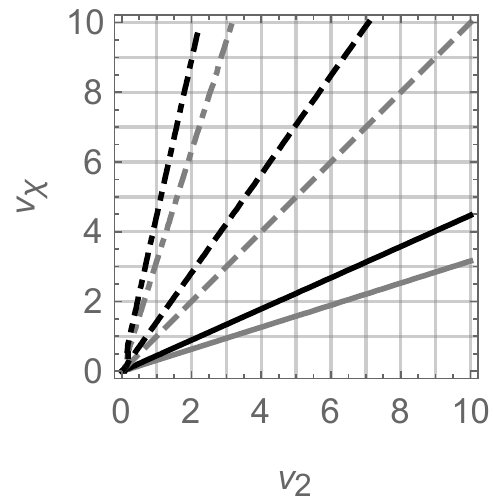}}
\subfigure[NO: ${h_{N\chi 1}}^{2} = 1.00$.]{\includegraphics[width=40mm]{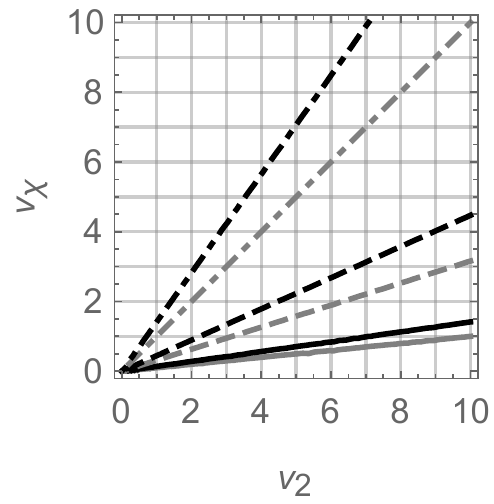}}
\caption{Contour plots of $v_{\chi}$ vs. $v_{2}$ from eq. \eqref{eq:Majorana-mass-NO-vChi-v2} for different values of ${h_{N\chi 1}}^{2}$ and $\mu_{N}$. From below to above there are the corresponding contour plots for the following values of $\mu_{N}$: 500 eV (gray, line), 1 keV (black, line), 5 keV (gray, dashed), 10 keV (black, dashed), 50 keV (gray, dot-dashed) and 100 keV (black, dot-dashed).} 
\label{fig:external-factor-NO}
\end{figure}

If the following constraint is assumed
\begin{equation}
\label{eq:Majorana-mass-NO-vChi-v2}
\frac{\mu_{N} v_{2}^{2}}{{h_{N\chi 1}}^{2}{v_{\chi}}^{2}} = 50\mathrm{\,meV},
\end{equation}
contour plots can be done for different values of $\mu_{N}$ in the $v_{\chi}$ vs. $v_{2}$ plane, as shown in figure \ref{fig:external-factor-NO}. 

\begin{table}
\centering
\begin{tabular}{c c c c c }
\hline\hline
			&	$\alpha_{e\mu}=0^{\mathrm{o}}$
			&	$\alpha_{e\mu}=15^{\mathrm{o}}$
			&	$\alpha_{e\mu}=30^{\mathrm{o}}$	\\\hline\hline
    $h_{2e}^{\nu e}$	&	$0.264 \rightarrow 0.278$
			&	$0.285 \rightarrow 0.299$
			&	$0.237 \rightarrow 0.270$	\\\hline
    $h_{2e}^{\nu \mu}$	&	$-0.707 \rightarrow -0.244$
			&	$-0.726 \rightarrow -0.335$
			&	$-0.796 \rightarrow -0.547$	\\
    $h_{2\mu}^{\nu \mu}$&	$-0.491 \rightarrow -0.190$
			&	$-0,464 \rightarrow -0.173$
			&	$-0.342 \rightarrow -0.039$	\\\hline
    $h_{2e}^{\nu \tau}$	&	$0.267 \rightarrow 0.748$
			&	$0.313 \rightarrow 0.677$
			&	$0.140 \rightarrow 0.355$	\\
    $h_{2\mu}^{\nu \tau}$&	$0.130 \rightarrow 0.462$
			&	$0.196 \rightarrow 0.460$
			&	$0.440 \rightarrow 0.510$	\\\hline\hline
\end{tabular}
\caption{Yukawa coupling domain which fulfil at $1\sigma$ neutrino oscillation data for NO reported by \cite{neutrinodata}. $h_{2\mu}^{\nu e}=0$ for simplifying the MonteCarlo search.}
\label{tab:Neutrino-parameters-NO}
\end{table}

\begin{table}
\centering
\begin{tabular}{c c c c c }
\hline\hline
			&	$\alpha_{e\mu}=0^{\mathrm{o}}$
			&	$\alpha_{e\mu}=1^{\mathrm{o}}$
			&	$\alpha_{e\mu}=2^{\mathrm{o}}$	\\\hline\hline
    $h_{2e}^{\nu e}$	&	$1.094 \rightarrow 1.107$
			&	$1.091 \rightarrow 1.105$
			&	$1.090 \rightarrow 1.103$	\\\hline
    $h_{2e}^{\nu \mu}$	&	$-0.122 \rightarrow -0.106$
			&	$-0.127 \rightarrow -0.113$
			&	$-0.128 \rightarrow -0.118$	\\
    $h_{2\mu}^{\nu \mu}$&	$0.970 \rightarrow 1.060$
			&	$0.980 \rightarrow 1.070$
			&	$1.010 \rightarrow 1.080$	\\\hline
    $h_{2e}^{\nu \tau}$	&	$0.110 \rightarrow 0.127$
			&	$0.122 \rightarrow 0.138$
			&	$0.135 \rightarrow 0.149$	\\
    $h_{2\mu}^{\nu \tau}$&	$0.930 \rightarrow 1.030$
			&	$0.920 \rightarrow 1.010$
			&	$0.910 \rightarrow 0.980$	\\\hline\hline
\end{tabular}
\caption{Yukawa coupling domain which fulfil at $1\sigma$ neutrino oscillation data for IO reported by \cite{neutrinodata}. $h_{2\mu}^{\nu e}=0$ for simplifying the MonteCarlo search.}
\label{tab:Neutrino-parameters-IO}
\end{table}

The tables \ref{tab:Neutrino-parameters-NO} and \ref{tab:Neutrino-parameters-IO} show regions where the neutrino Yukawa couplings and the angle $\alpha_{e\mu}$ make consistent this model with neutrino oscillation data reported by \cite{neutrinodata} at $3\sigma$. 

The Yukawa coupling $h_{\chi N 3}$ is not fixed by oscillations of the light neutrinos; however, they may contribute into the total rotation matrix $\mathbb{U}_{N}$ in \eqref{eq:Total-neutrino-mixing-matrix}. Thus, the neutral spectrum of the model is composed by three active light neutrinos $\nu^{1,2,3}_{L}$ and six quasidegenerated steril neutrinos $N^{1,2,3}_{1L}$ and $N^{1,2,3}_{2L}$ at the TeV scale. 

\section{Conclusions}
\label{sect:Conclusions}
Abelian nonuniversal gauge extensions of the SM are very well-motivated models which involve a wide number of theoretical aspects. In this work, by requiring nonuniversality in the left-handed quark sector and in lepton sector, we propose a new $G_{SM} \times \mathrm{U}(1)'$ gauge model. We obtained a free-anomaly theory with invariant Yukawa interactions, predicting hierarchical mass structures in the quark and charged lepton sector with few free parameters 

For the quark sector, we identify three energy scales. First, at the breaking scale of the $U(1)_X$ symmetry, we obtain heavy masses to the extra heavy quarks $J^{n}$ and $T$, with $M_{J^{n}}\approx M_{T} \sim \upsilon _X$. Second, at tree level, we obtain masses at the electroweak scale for the $c$, $t$ and $b$ quarks, with $M_{c,t,u} \sim \upsilon _{1,2}$. Finally, at one-loop level, we obtain light masses for the $u$, $d$ and $s$ quarks, with $M_{u,d,s} \sim \upsilon _{1,2}^2/\upsilon _{\chi}$. For the leptonic sector, we also obtain the same hierarchical structure, where the extra leptons $E$ and $\mathcal{E}$ acquire masses at the $\upsilon _{\chi}$ scale, the $\mu$ and $\tau$ have masses at the electroweak scale, and the electron obtain masses at one-loop, which is suppressed as $\upsilon _{1,2}^2/\upsilon _{\chi}$ .
 
On the other hand, with the addition of extra Majorana neutrinos, we found that neutrinos may acquire tiny masses via the inverse seesaw mechanism. The selection of a small Majorana mass term (from eV to KeV scale) and the experimental limits on observables from neutrino oscillations allows us to perform numerical adjustment  for the values of the Yukawa couplings of neutrinos in NO and IO scenarios. In addition, because the non-universal $\mathrm{U(1)}_{X}$ charges, the electron remains massless at tree level but a non-vanishing mass term emerges at one-loop corrections which gives a viable explanation for its small mass compared to the electroweak scale.

\section*{Acknowledgment}
This work was supported by \textit{El Patrimonio Autonomo Fondo Nacional de Financiamiento para la Ciencia, la Tecnolog\'{i}a y la Innovaci\'on Francisco Jos\'e de Caldas} programme of COLCIENCIAS in Colombia. RM thanks to professor Germán Valencia for the kindly hospitality at Monash University and his useful comments.

\appendix

\section{Block Diagonalization}\label{app:block}

Let us take a generic matrix with arbitrary dimension of the form:
\begin{eqnarray}
\mathbb{M}^2=
\begin{pmatrix}
A & C \\
C^T & D
\end{pmatrix},
\label{block-matrix}
\end{eqnarray}
with $A, D$ and $C$ sub-matrices whose elements obey the hierarchy 
\begin{eqnarray}
A \ll C \ll D.
\label{block-hierarchy}
\end{eqnarray} 
The matrix (\ref{block-matrix}), as shown in reference \cite{grimus}, can be block diagonalized approximately by a unitary rotation of the form:
\begin{eqnarray}
V=
\begin{pmatrix}
I & F \\
-F^T & I
\end{pmatrix},
\label{approx-rotation}
\end{eqnarray}
where $I$ is an identity matrix, and $F$ a small sub-rotation with $F\ll 1$. Keeping only up to linear terms on $F$, the rotation gives:
\begin{eqnarray}
V^T \mathbb{M}^2 V=
\begin{pmatrix}
A-CF^T-FC^T & C+AF-FD \\
C^T+F^TA-DF^T & D+C^TF+F^TC
\end{pmatrix},
\label{rotation}
\end{eqnarray}
which, by definition, must lead us to a diagonal block form
\begin{eqnarray}
\mathbbm{m}^2=
\begin{pmatrix}
a & 0 \\
0 & d
\end{pmatrix},
\label{diagonal-block}
\end{eqnarray}
with $a$ and $d$ non-diagonal matrices, and $0$ the null matrix. By matching the upper right non-diagonal block in (\ref{rotation}) and (\ref{diagonal-block}), we obtain that $C+AF-FD=0$. Taking into account the hierarchy in (\ref{block-hierarchy}), we may neglect the term with $A$, finding the following approximate solution:
\begin{eqnarray}
F\approx CD^{-1}.
\label{F-subrotation}
\end{eqnarray}

On the other hand, if we match the diagonal blocks in (\ref{rotation}) and (\ref{diagonal-block}), and using the solution (\ref{F-subrotation}), we can obtain the form of the submatrices $a$ and $b$ in terms of the original blocks $A$, $C$ and $D$. We obtain at dominant order that:
\begin{eqnarray}
a &\approx & A-CD^{-1}C^T \nonumber \\
b &\approx & D.
\end{eqnarray}
The above matrices can be diagonalized independently.

\end{document}